\documentclass[12pt,a4paper]{article}
\pdfoutput=1
\usepackage[affil-sl,auth-sc]{authblk}
\usepackage{amsfonts,amsmath,amssymb}
\usepackage{gensymb}
\usepackage{cite}
\usepackage{mathrsfs,graphicx}
\usepackage{bm}
\usepackage{color}
\usepackage[hyperfootnotes=false,linktocpage=true]{hyperref}
\usepackage{url}
\usepackage[utf8]{inputenc}
\usepackage[nottoc]{tocbibind} 

\evensidemargin=\oddsidemargin

\newcommand{\bea}{\begin{eqnarray}}
\newcommand{\eea}{\end{eqnarray}}
\def\beq{\begin{equation}}
\def\eeq{\end{equation}}

\newcommand{\nn}{\nonumber\\}

\newcommand{\gsim}{\lower.7ex\hbox{$\;\stackrel{\textstyle>}{\sim}\;$}}
\newcommand{\lsim}{\lower.7ex\hbox{$\;\stackrel{\textstyle<}{\sim}\;$}}

\def\nnb{\nonumber}

\numberwithin{equation}{section}

\begin{document}

\title{
Analytical description of CP violation in oscillations of atmospheric
neutrinos traversing the Earth
}

\author[1,2]{Ara Ioannisian
\thanks{ara.ioannisyan@cern.ch}
}

\author[3]{Stefan Pokorski
\thanks{stefan.pokorski@fuw.edu.pl}
}

\author[3]{Janusz Rosiek
\thanks{janusz.rosiek@fuw.edu.pl}
}

\author[3]{Micha{\l} Ryczkowski}

\affil[1]{Yerevan Physics Institute, Alikhanian Br. 2, 375036 Yerevan,
  Armenia }

\affil[2]{Institute for Theoretical Physics and Modelling, 375036
  Yerevan, Armenia }
\affil[3]{Faculty of Physics, University of Warsaw, Pasteura
  5,\protect\linebreak 02-093 Warsaw, Poland}

\date{June 24, 2020}
  
\maketitle


\begin{abstract}
Flavour oscillations of sub-GeV atmospheric neutrinos and
antineutrinos, traversing different distances inside the Earth, are a
promising source of information on the leptonic CP phase $\delta$.  In
that energy range, the oscillations are very fast, far beyond the
resolution of modern neutrino detectors. However, the necessary
averaging over the experimentally typical energy and azimuthal angle
bins does not wash out the CP violation effects. In this paper we
derive very accurate analytic compact expressions for the averaged
oscillations probabilities.  Assuming spherically symmetric Earth, the
averaged oscillation probabilities are described in terms of two
analytically calculable effective parameters.  Based on those
expressions, we estimate maximal magnitude of CP-violation effects in
such measurements and propose optimal observables best suited to
determine the value of the CP phase in the PMNS mixing matrix.
%
%
\end{abstract}

\newpage

\tableofcontents

\newpage


\section{Introduction}
\label{sec:intro}

Determination of the leptonic CP phase by measuring neutrino
oscillations is a challenging issue~\cite{Donini:1999, Ohlsson:1999um,
  Farzan:2002ct, Nunokawa:2007, Akhmedov:2008, Branco:2012,
  Ohlsson:2013, Razzaque:2014, Machado:2014, Bernabeu:2018,
  Kelly:2019itm}.  It is well known that sensitivity of oscillations
to the CP phase $\delta$ generically decreases with the increasing
neutrino energy.  Matter effects may be helpful in measuring $\delta$
but they also fade away when the neutrino energy
increases~\cite{Barger:1980}.  Thus the oscillations of low energy
atmospheric neutrinos, hitting a detector at different angles, after
traversing different distance inside the Earth, look as a particularly
promising source of information on the leptonic
$\delta$~\cite{Kelly:2019itm,IoaDune:2018}.  However, in that case the
limitations come from the difficulties with precise determination of
the neutrino energy and the angle it hits a detector.  Therefore,
analytical understanding of the oscillation probabilities for low
energy (say, below ${\cal O}(1)$ GeV) atmospheric neutrinos, as a
function of their energy and the number of layers they traverse in the
Earth, would be very useful for optimising measurements of the
leptonic CP phase in realistic experimental setups.  This is the
purpose of the present paper.  Oscillations of sub-GeV neutrinos
differ substantially from those of higher energy neutrinos~\cite{
  Akhmedov:2008, Barger:1998, Peres:2004, Friedland:2004, Huber:2005,
  Hay:2012, Agarwalla:2012, Blennov:2013}.  First of all, they are
very fast in energy, far beyond the energy resolution of modern
neutrino detectors, \cite{DUNE, HYPERK}, because they are affected by
both solar and atmospheric mass splittings.  Thus, the relevant
``observables'' carrying the physical information are the oscillation
probabilities averaged over typical experimental energy and angle
bins.  In addition, the patterns of matter effects also change with
energy~\cite{Wolfenstein, Smirnov:1985, Akhmedov:1988}.  A useful
insight can be obtained from the description of oscillation
probabilities in matter in the conventional parametric form as in the
vacuum but with effective mixing angles and mass
eigenvalues~\cite{Ioannisian:2018qwl,
Wang:2019yfp, Wang:2019dal}.  The main effect resides in the energy
dependence of effective mixing angles $\theta^m_{12}$ and
$\theta^m_{13}$.  At sub-GeV energies and the matter densities typical
for the Earth structure, $\theta^m_{13}$ is close to, and
$\theta^m_{12}$ is significantly different from their vacuum values,
whereas the opposite is true at higher energies (in that
parametrization $\theta^m_{23}$ remains to be the vacuum angle).

In this paper we derive analytical parametrization of the averaged
oscillation probabilities for sub-GeV neutrinos, after traversing
arbitrary number of Earth layers, each with a constant matter density.
For the spherically symmetric Earth, which is a very good
approximation once the fast oscillations are averaged out, the
oscillation probabilities are described in terms of two effective
parameters.  Based on those expressions, we estimate maximal magnitude
of CP-violation effects in such measurements.  We also propose optimal
observables best suited to determine the value of the CP-phase in the
PMNS mixing matrix.

Our article is organised as follows. In Section~\ref{sec:exact} we
derive the exact formulae for the transition matrix for neutrinos
traversing the Earth, divided into layers of constant matter density.
In Section~\ref{sec:papprox} we propose the approximations which can
be done for the considered neutrino energy range and symmetric Earth
layout and we derive simple and accurate analytical formulae for the
averaged oscillation probabilities.  Section~\ref{sec:genopti} is
devoted to the discussion of the optimal experimental setup and choice
of observables best suited to measure the leptonic CP-phase.  In
Section~\ref{sec:finbin} we discuss the dependence of the averaged
oscillation probabilities on the size of experimental bins in energy
and azimuthal angle.  We conclude in Section~\ref{sec:summary}. In
Appendix~\ref{app:trig} for completeness we collect the formulae for
the neutrino track lengths in the Earth layers.  Finally in
Appendix~\ref{app:quality} we discuss the numerical quality of the
approximations done when deriving the analytical formulae.

\section{Oscillation probabilities for neutrinos traversing the Earth}.
\label{sec:exact}

We consider neutrino oscillations when traversing the Earth.  Our main
focus is on sub-GeV atmospheric neutrinos but the framework we develop
in this Section is a general one.  In the next Section we shall
discuss the approximations appropriate for low energy neutrinos.

\begin{figure}[tb!]
\begin{center}
 \includegraphics[width=0.6\textwidth]{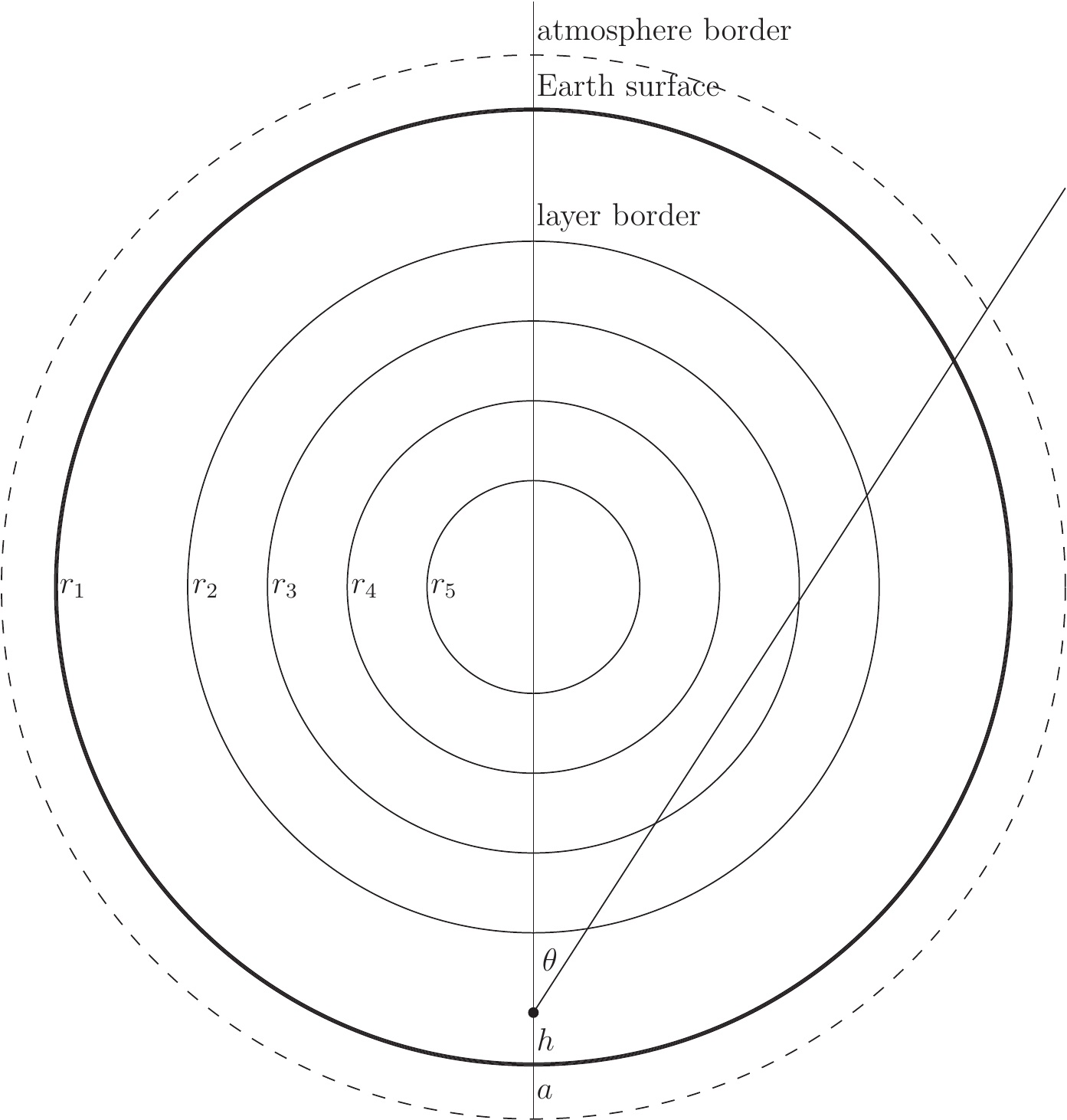}
\end{center}

\caption{Schematic picture of the Earth structure (not to scale) and
  definition of the azimuthal angle $\theta$.  The detector, marked by
  black blob, is located at depth $h$ below the Earth surface, the
  average atmosphere width is denoted by $a$.\label{fig:earth}}
\end{figure}

In order to estimate possible effects of the CP phase in the PMNS
mixing matrix on the transition probabilities, we calculate them
analytically assuming the Earth structure based on the PREM
model~\cite{PREM}.  In such an approximation the Earth is divided into
a finite number of layers, each having a constant density of matter.
Although our analysis can be applied to any number of layers, for
numerical estimates we use 5-layer pattern of the structure of our
planet - starting from the center, one has inner core, outer core,
lower mantle, upper mantle and crust.  Our schematic setup is
illustrated in Fig.~\ref{fig:earth}, and the layer radii and densities
are collected in Table~\ref{tab:earth}.  Depending on the azimuthal
angle $\theta$, neutrinos can traverse 1,3,5,7 or 9 Earth layers.
Other numerical inputs used throughout the paper are collected in
Table~\ref{tab:input}.

\begin{table}[htb!]
\begin{center}
  \begin{tabular}{|c|c|c|c|}
    \hline
Layer number & External radius & Density (Avogadro units) & Neutrino
potential (MeV) \\
    \hline
    1 & 1 &    1.69 & $1.29 \cdot 10^{-19}$ \\
    2 & 0.937 &  1.92 & $1.47 \cdot 10^{-19}$\\
    3 & 0.895 & 2.47 & $1.88 \cdot 10^{-19}$\\
    4 & 0.546 & 5.24 & $4.00 \cdot 10^{-19}$\\
    5 & 0.192 & 6.05 & $4.63 \cdot 10^{-19}$\\
    \hline
\end{tabular}
\end{center}

\caption{External layer radii as a fraction of the Earth radius
  $R=6371$ km, average layer densities and corresponding neutrino
  interaction potential.\label{tab:earth}}

\end{table}

\begin{table}[htb!]
\begin{center}
  \begin{tabular}{|c|c|c|}
    \hline
    Quantity       & Value (NO) & Value (IO)   \\
    \hline
    & & \\[-1mm]
$\Delta m_{a}^2$ & $ (2.50\pm 0.03)\cdot 10^{-15}$ MeV$^2$ &$
    - (2.42^{+0.03}_{-0.04})\cdot 10^{-15}$ MeV$^2$ \\[2mm]
$\Delta m_\odot^2$ & $(7.55^{+0.20}_{-0.16})\cdot 10^{-17}$ MeV$^2$ &
    $(7.55^{+0.20}_{-0.16})\cdot 10^{-17}$ MeV$^2$\\[2mm]
$\theta_{12}$ & $(34.5^{+1.2}_{-1.0})^{\degree}$ &
    $(34.5^{+1.2}_{-1.0})^{\degree}$ \\[2mm]
$\theta_{23}$ & $(47.7^{+1.2}_{-1.7})^{\degree}$ &
    $(47.9^{+1.0}_{-1.7})^{\degree}$ \\[2mm]
$\theta_{13}$ & $(8.45^{+0.16}_{-0.14})^{\degree}$ &
    $(8.53^{+0.14}_{-0.15})^{\degree}$\\[2mm]
\hline
\end{tabular}
\end{center}

\caption{Neutrino mass differences and mixing angles in the vacuum
  used throughout the paper for normal mass ordering (NO) and inverted
  ordering (IO)~\cite{deSalas:2017kay}.  We denote $\Delta
  m_{a}^2=m^2_3-m^2_1$ and $\Delta m_\odot^2 =
  m^2_2-m^2_1$.  \label{tab:input}}

\end{table}

The neutrino oscillation probabilities are determined by the
$S$-matrix elements ($\alpha,\beta = e,\mu,\tau$):
\bea
S_{\alpha \beta} =T \ e^{-i \int_{x_0}^{x_f} {\cal H}(x) dx}
\eea
with
\bea
{\cal H}=U \left( \begin {array}{ccc} 0 & 0 & 0\\
0 &{\Delta m^2_\odot \over 2E}&0\\
0 & 0 &{\Delta m^2_a \over 2E} \end {array} \right)
U^\dagger + \left( \begin {array}{ccc} V(x) & 0 & 0\\ 0 & 0 & 0 \\ 0 &
  0 & 0 \end {array} \right)
\label{eq:electroweak}
\eea
where $U$ denotes the neutrino mixing matrix in the vacuum, with the
parametrization:
\bea
U = O_{23} \, U_\delta\, O_{13}\, O_{12}
\eea
and
\bea
O_{12} &=& \left( \begin {array}{ccc} \cos\theta_{12} &
  \sin\theta_{12} & 0\\
-\sin\theta_{12} & \cos\theta_{12} & 0\\
0 & 0 & 1 \end {array} \right)
\qquad
O_{13} = \left( \begin {array}{ccc} \cos\theta_{13} & 0 &
  \sin\theta_{13} \\
0 & 1 & 0\\
-\sin\theta_{13} & 0 & \cos\theta_{13} \end {array} \right)
\nnb\\[3mm]
O_{23} &=& \left( \begin {array}{ccc} 1 & 0 & 0 \\
0 & \cos\theta_{23} & \sin\theta_{23} \\
0 & -\sin\theta_{23} & \cos\theta_{23} \end {array} \right)
\qquad
U_\delta = \left( \begin {array}{ccc} 1 & 0 & 0 \\
0 & 1 & 0\\
0 & 0 & e^{i\delta} \end {array} \right)
\eea
The $V(x)$ is the neutrino weak interaction potential energy in
matter.  As shown in ref.~\cite{Ioannisian:2018qwl}, for a constant
$V$, to a good approximation the Hamiltonian ${\cal H}$ can be
diagonalized by the matrix $U_m$ where only the $\theta_{12}$ and
$\theta_{13}$ angles have values modified compared to vacuum:
\bea
{\cal H} = U_m \left( \begin {array}{ccc} {\cal H}_1 & 0 & 0\\
0 & {\cal H}_2 & 0\\
0 & 0 & {\cal H}_3 \end {array} \right) U_m^\dagger
& \equiv & U_m {\cal H}_d U_m^\dagger
\eea
with
\bea
U_m = O_{23} \, U_\delta\, O_{13}^m\, O_{12}^m
\label{eq:um}
\eea
and approximate explicit formulae for $O_{13}^m, O_{12}^m$ are given
in ref.~\cite{Ioannisian:2018qwl}.

It is convenient to work in a new basis, rotated by the matrix
\bea
U_0 = O_{23} \, U_\delta\, O_{13}
\eea
so that the rotated Hamiltonian has the form 
\bea
{\cal H'} \equiv U_0^\dagger {\cal H} U_0 = \left( \begin {array}{ccc}
  {\Delta m^2_\odot \over 2E} s_{12}^2 & {\Delta m^2_\odot \over 2E}
  s_{12} c_{12} & 0\\
{\Delta m^2_\odot \over 2E} s_{12} c_{12} & {\Delta m^2_\odot \over
  2E} c_{12}^2& 0\\
0 & 0 &{\Delta m^2_a \over 2E} \end {array} \right)
 + V \left( \begin {array}{ccc} c_{13}^2 & 0 & s_{13}
  c_{13} \\ 0 & 0 & 0 \\ s_{13} c_{13} & 0 & s_{13}^2 \end {array}
\right)
\label{eq:rotbasis}
\eea
and can be approximately diagonalized by $12, 13$ rotations only, with
angles including matter effects:
\bea
{\cal H}' = O_{13}^T \, O^m_{13}\, O^m_{12} {\cal H}_d \, O^{mT}_{12}
\, O^{mT}_{13} \, O_{13}
\eea

The transition matrix $S$ can written as the time-ordered product of
the transition matrices in the Earth layers,
\bea
S = T\, \Pi_i S_i
\label{eq:stprod}
\eea
where within the $i$-th layer of constant density the matrix $S_i$ is
simply given by
\bea
S_i = e^{-i {\cal H}_i \Delta x_i}
\eea
Using the rotated basis defined above, one can easily show that (up to
an unimportant overall phase denoted as $e^{i\xi}$) the matrix $S$ can
be expressed as
\bea
S = U_0 \left( T\, \Pi_i e^{-i {\cal H}'_i \Delta x_i} \right)
U_0^\dagger = e^{i\xi} U_a \, T\, \Pi_i \left( O^m_{i13}\, O^m_{i12}
{\cal E}_i \, O^{mT}_{i12} \, O^{mT}_{i13} \right) \, U_a^\dagger
\label{eq:numstart}
\eea
where we have defined
\bea
U_a &=& O_{23} \, U_\delta\nnb\\[3mm]
{\cal E}_i &=& \left( \begin {array}{ccc}
e^{\frac{1}{2} i({\cal H}^i_2-{\cal H}^i_1)\Delta x_i} & 0 & 0\\
0 & e^{-\frac{1}{2}i({\cal H}^i_2-{\cal H}^i_1)\Delta x_i} & 0\\
0 & 0 & e^{-i({\cal H}^i_3-\frac{{\cal H}^i_1 + {\cal H}^i_2}{2}
  )\Delta x_i}
\end{array} \right)
\label{eq:ei}
\eea
Formula~(\ref{eq:numstart}) is general and does not involve any
approximation yet (other than the ``layered Earth'' model).  In the
next Section we introduce analytical approximations appropriate for
the oscillation probabilities of the sub-GeV atmospheric neutrinos.

\section{Analytical approximations for sub-GeV atmospheric neutrinos}
\label{sec:papprox}

\subsection{Averaging of probabilities over energy bins}
\label{sec:paver}

The transition probabilities for sub-GeV atmospheric neutrinos
oscillate quickly with neutrino energy and azimuthal angle.  This can
be traced back to the fact that small variations of both quantities
can significantly change the ratio $\Delta m^2_a L(\theta)/E$.
Realistically, the oscillation probabilities have to be averaged over
bins in energy and angle corresponding to the relevant experimental
resolutions. As long as the period of the neutrino oscillation
frequency is far smaller than the experimental resolution significant
simplifications can be performed in calculating analytically the
averaged oscillation probabilities.

\begin{figure}[tb!]
\begin{center}
\includegraphics[width=0.7\textwidth]{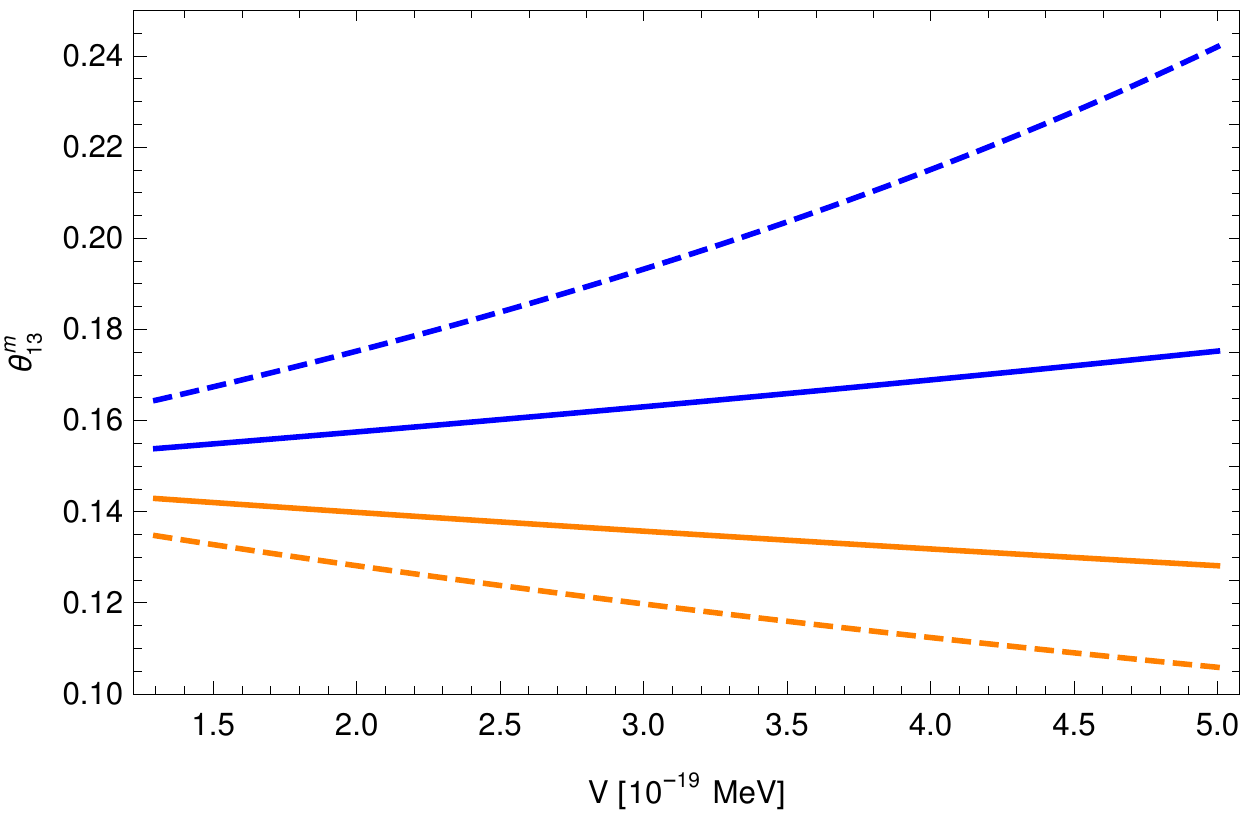}
\end{center}

\caption{$\theta_{13}^m$ in matter as the function of neutrino
  interaction potential in Earth for normal (blue line) and inverted
  (orange line) mass ordering and for the neutrino energies $E=400$
  MeV (solid lines) and $E=1000$ MeV (dashed lines).
 \label{fig:angleveff13}}

\end{figure}

First, we observe that in the product~(\ref{eq:numstart}) the
following structure repeats itself:
\bea
\ldots {\cal E}_i \, O^{mT}_{i12} \, O^{mT}_{i13} \, O^m_{(i+1)13}\,
O^m_{(i+1)12} \,{\cal E}_{i+1} \ldots
\eea
with the most inner multiplication matrix depending on the differences
of the $\theta_{13}^m$ mixing angle between the neighbouring layers:
\bea
O^{mT}_{i13} \, O^m_{(i+1)13} = \left( \begin {array}{ccc}
  \cos(\theta_{i13}^m - \theta_{(i+1)13}^m) & 0 & \sin (\theta_{i13}^m
  - \theta_{(i+1)13}^m) \\
0 & 1 & 0\\
-\sin (\theta_{i13}^m - \theta_{(i+1)13}^m) & 0 & \cos (\theta_{i13}^m
- \theta_{(i+1)13}^m) \end {array} \right)
\label{eq:th13jump}
\eea
Contrary to high energy neutrinos, like for instance in the Dune
experiment~\cite{Ioannisian:2018qwl}, for the neutrino energies below
$E<{\cal O}(1)$ GeV and typical values of the Earth density,
$\theta_{i13}^m$ angle in matter vary only very slightly, as
illustrated in Fig.~\ref{fig:angleveff13}.  The differences
$\theta_{i13}^m - \theta_{(i+1)13}^m$ between the layers are typically
of the order of $0.01$ radian, even less for the lower neutrino
energies.  Therefore, to a good approximation products of
$O^{mT}_{i13} \, O^m_{(i+1)13}$ can be replaced by the unit matrices.
Note, in particular, that for $E<{\cal O}(1)$ GeV and the matter
densities in the Earth layers in the range
$(3-12)\frac{\mathrm{g}}{\mathrm{cm}^3}$ we are well below the
resonantly enhanced values of $\theta^m_{13}$.  In contrast, the
dependence of the $\theta_{i12}^m$ on the matter density is stronger
for this energy range.

Then, neglecting the overall phase, the time-ordered product on the
RHS of eq.~(\ref{eq:numstart}) takes the form
\bea
O^m_{13-first} \, T\, \Pi_i \left( O^m_{i12} {\cal E}_i \,
O^{mT}_{i12} \, \right) O^{mT}_{13-last}
\label{eq:prod2}
\eea
First layer on the neutrino track is the atmosphere, so that
$O^m_{13-first} \approx O_{13}$ in vacuum.  The last layer is the
Earth crust around the detector, so that $O^{m}_{13-last} =
O^{m}_{13-crust}$.  The inner product in eq.~(\ref{eq:prod2}) contain
only $O_{i12}^m$ mixing matrices, thus the result has the structure:
\bea
T\, \Pi_i \left( O^m_{i12} {\cal E}_i \, O^{mT}_{i12} \, \right)
= \left( \begin {array}{ccc} X_{11} & X_{12} & 0 \\
X_{12} & X_{22} & 0\\
0 & 0 & 0 \end {array} \right)
+ \left( \begin {array}{ccc} 0 &
  0 & 0 \\
0 & 0 & 0\\
0 & 0 & \Pi_i ({\cal E}_i)_{33} \end {array} \right)
\label{eq:prod3}
\eea
Approximating again the product $O^{mT}_{13-crust} O_{13}$ by the unit
matrix we arrive at the following expression for the transition matrix
$S$ (up to an unimportant overall phase factor):
\bea
S &\approx&  U_0 
\left(\begin {array}{ccc} X_{11} & X_{12} & 0 \\
X_{12} & X_{22} & 0\\
0 & 0 & 0 \end {array} \right) U_0^\dagger
+ \Pi_i ({\cal E}_i)_{33} \, U_0 
\left( \begin{array}{ccc} 0 & 0 & 0 \\
0 & 0 & 0\\
0 & 0 & 1 \end {array} \right) 
U_0^\dagger\nnb\\[4mm]
&\equiv& A + \Pi_i ({\cal E}_i)_{33}\, B
\label{eq:sfin}
\eea
The matrix $A=A(E,\theta)$ can be calculated by the numerical
diagonalization of the Hamiltonian given in eq.~(\ref{eq:electroweak})
and the matrix $B$ is constant, given by the vacuum mixing angles.
Oscillation probabilities are given by
\bea
P_{\alpha\beta} = |S_{\beta\alpha}|^2 = |A_{\beta\alpha}|^2 + 2
\mathrm{Re}\left[ A^*_{\beta\alpha} B_{\beta\alpha} \, \Pi_i({\cal
    E}_i)_{33} \right] + |B_{\beta\alpha}|^2
\label{eq:pnonaver}
\eea
In eq.~(\ref{eq:pnonaver}) only the quantity $\Pi_i ({\cal E}_i)_{33}$
(being a pure phase) depends on the larger neutrino mass splitting and
varies quickly with energy and azimuthal angle.  When averaged over
bins in energy $\Delta E$ and angle $\Delta\theta$ {\it larger than
  the period of the oscillation frequency}, that term vanishes and we
get:
\bea
\bar P_{\alpha\beta}(E,\theta) &=& \frac{1}{\Delta E \Delta \theta}
\int_{E - \frac{\Delta E}{2}}^{E + \frac{\Delta E}{2}} \int_{\theta -
  \frac{\Delta \theta}{2}}^{\theta + \frac{\Delta \theta}{2}}
P_{\alpha\beta}(E',\theta') dE'd\theta'\nnb\\[2mm]
&=&\frac{1}{\Delta E \Delta \theta} \int_{E - \frac{\Delta E}{2}}^{E +
  \frac{\Delta E}{2}} \int_{\theta - \frac{\Delta \theta}{2}}^{\theta
  + \frac{\Delta \theta}{2}} |A_{\beta\alpha}|^2 dE'd\theta' +
|B_{\beta\alpha}|^2
\label{eq:PAB}
\eea
This is the first important result of the paper -- the averaging over
energy and azimuthal angle can be now done using some standard
2-dimensional numerical integration techniques, expected to be quickly
converging and accurate as the numerically most difficult and CPU-time
consuming averaging over fast oscillations of probabilities has been
done analytically while obtaining the formulae ~(\ref{eq:PAB}).

Furthermore, as we show below, one can also derive for the matrix $A$,
and thus for the integrand in eq.~(\ref{eq:PAB}), an excellent
analytical approximation in terms of only two effective parameters.
For neutrino energies larger than $300-400$ MeV, they are very
accurately calculable analytically.  Clearly, formula~(\ref{eq:PAB})
is useful when typical experimental bins in energy and azimuthal angle
are bigger than the period of oscillation frequencies.  As discussed
in the Appendix~\ref{app:paver} this is true for sub-GeV neutrino
energies. For higher energies $E\gsim 1$ GeV the probabilities defined
in eq.~(\ref{eq:PAB}) do not agree well with the
formulae~(\ref{eq:pnonaver}).

\subsection{Analytical results for the matrix $A$}
\label{sec:xprop}

We begin with the discussion of the properties of the matrix $X$.  The
full $2\times 2$ matrix $X$ defined in eq.~(\ref{eq:prod2}) is a
time-ordered product of matrices of the form $X_i = O^m_{i12} {\cal
  E}_i \, O^{mT}_{i12}$ (one for each Earth layer).  With the phase
conventions chosen in eq.~(\ref{eq:ei}), each of matrices $X_i$ is
unitary, symmetric and have determinant equal to 1.  Any such matrix
has only 2 free real parameters and can be expressed as:
\bea
X_i(\alpha_i,\phi_i) = \left(\begin{array}{cc}
\cos\alpha_{i}\, e^{-i\phi_{i}} & -i\sin\alpha_{i} \\
-i\sin\alpha_{i} & \cos\alpha_{i} \, e^{i\phi_{i}}
\label{eq:xidef}
\end{array}\right)
\eea
Defining 
\bea
\nu_i = ({\cal H}^i_2-{\cal H}^i_1)\Delta x_i
\label{eq:nudef}
\eea
direct calculations lead to the formulae
\bea
X_i = \left(\begin{array}{cc}
\cos\frac{\nu_i}{2} + i \cos 2\theta_{i12}^m \sin\frac{\nu_i}{2} & -i
\sin 2\theta_{i12}^m \sin\frac{\nu_i}{2} \\[4mm]
-i \sin 2\theta_{i12}^m \sin\frac{\nu_i}{2} & \cos\frac{\nu_i}{2} - i
\cos 2\theta_{i12}^m \sin\frac{\nu_i}{2}
\end{array}\right)
\label{eq:xnudef}
\eea
so that comparing with eq.~(\ref{eq:xidef}) one has
\bea
\sin\alpha_i &=& \sin 2\theta_{i12}^{m}\, \sin\frac{\nu_i}{2}\nnb\\
\tan\phi_i &=& -\cos 2\theta_{i12}^m \tan\frac{\nu_i}{2}
\label{eq:xangles}
\eea
Let us note that excluding azimuthal angles close to $\pi/2$ or bigger
(when the length of the neutrino track in the atmosphere and the
asymmetric position of the detector under the Earth surface cannot be
neglected) our setup is symmetric with respect to the Earth center.
Thus, the full matrix $X$ is to a good approximation given by a
symmetric product of $X_i$ and has the same symmetry properties as
each of them separately:
\bea
X = X_1 \ldots X_{k-1} X_k X_{k-1} \ldots X_1 \approx \left(
\begin{array}{cc}
\cos\alpha_X e^{-i\phi_X} &  - i \sin\alpha_X \\
- i \sin\alpha_X & \cos\alpha_X e^{i\phi_X}
\end{array}  
\right)
\label{eq:xfull}
\eea
The quality of this approximation turns out to be very good, as
discussed in Appendix~\ref{app:numphi}.

Using the parametrization of eq.~(\ref{eq:xfull}), one can derive
compact expressions in terms of the effective parameters $\phi_X,
\alpha_X$ for the $\nu_\alpha\to \nu_\beta$ oscillation probabilities
given by eq.~(\ref{eq:PAB}):
\bea
\bar P_{\alpha\beta}(E,\theta) = \frac{1}{\Delta E \Delta \theta}
\int_{E - \frac{\Delta E}{2}}^{E + \frac{\Delta E}{2}} \int_{\theta -
  \frac{\Delta \theta}{2}}^{\theta + \frac{\Delta
    \theta}{2}}I_{\alpha\beta}(E',\theta')dE' d\theta'
\label{eq:PI}
\eea
where the matrix elements of $I_{\alpha\beta}$ defined as
\bea
I_{\alpha\beta}  =  |A_{\beta\alpha}|^2 + |B_{\beta\alpha}|^2
\label{eq:idef}
\eea
are given by .
\bea
I_{ee} & = & \sin^4\theta_{13} + \cos^4\theta_{13}\cos^2\alpha_X\nnb\\[2mm]
I_{e\mu} &=& 2 \cos^2\theta_{13} \sin^2\theta_{13} \sin^2\theta_{23} +
\cos^2\theta_{13} (\cos^2\theta_{23} - \sin^2\theta_{13}
\sin^2\theta_{23} )\sin^2\alpha_X\nnb\\
&+&\frac{1}{2} \cos^2\theta_{13} \sin\theta_{13} \sin2 \theta_{23}
\sin2 \alpha_X \sin(\delta - \phi_X) \nnb\\[2mm]
I_{\mu e} &=& 2 \cos^2\theta_{13} \sin^2\theta_{13} \sin^2\theta_{23} +
\cos^2\theta_{13} ( \cos^2\theta_{23}- \sin^2\theta_{13}
\sin^2\theta_{23} )\sin^2\alpha_X \nnb\\
&-& \frac{1}{2} \cos^2\theta_{13} \sin\theta_{13} \sin2 \theta_{23}
\sin 2 \alpha_X \sin(\delta + \phi_X)\nnb\\[2mm]
I_{\mu\mu} &=& \cos^4\theta_{13} \sin^4\theta_{23}
 + \cos^2\alpha_X (\cos^4\theta_{23} + \sin\theta_{13}^4
\sin^4\theta_{23} + \frac{1}{2} \cos2 \phi_X \sin^2\theta_{13} \sin^2
2 \theta_{23}) \nnb\\
&+& \sin\theta_{13} (\cos^2\theta_{23} - \sin^2\theta_{13}
\sin^2\theta_{23}) \sin 2 \theta_{23} \sin 2 \alpha_X \sin\phi_X
\cos\delta \nnb\\
&+& \sin^2\theta_{13} \sin^2 2 \theta_{23} \sin^2\alpha_X \cos^2\delta
\label{eq:asolve}
\eea
The same formulae hold for the antineutrino oscillation probabilities,
after replacing $\delta\rightarrow -\delta$ and using effective
parameters $\bar\phi_X$, $\sin\bar\alpha_X$ describing antineutrino
mixing (see discussion below and eq.~(\ref{eq:antiphialexp})).

For narrow energy and azimuthal angle bins one has $\bar
P_{\alpha\beta}(E,\theta) \approx I_{\alpha\beta}(E,\theta)$.  In
Sec.~\ref{sec:genopti} we discuss qualitative properties of $\bar
P_{\alpha\beta}(E,\theta)$ using this approximation, i.e.  assuming
both quantities to be equivalent.  Averaging over wider energy and
azimuthal angle bins is discussed in more details in
Sec.~\ref{sec:finbin}.

\begin{figure}[tb!]
\begin{center}
\includegraphics[width=0.7\textwidth]{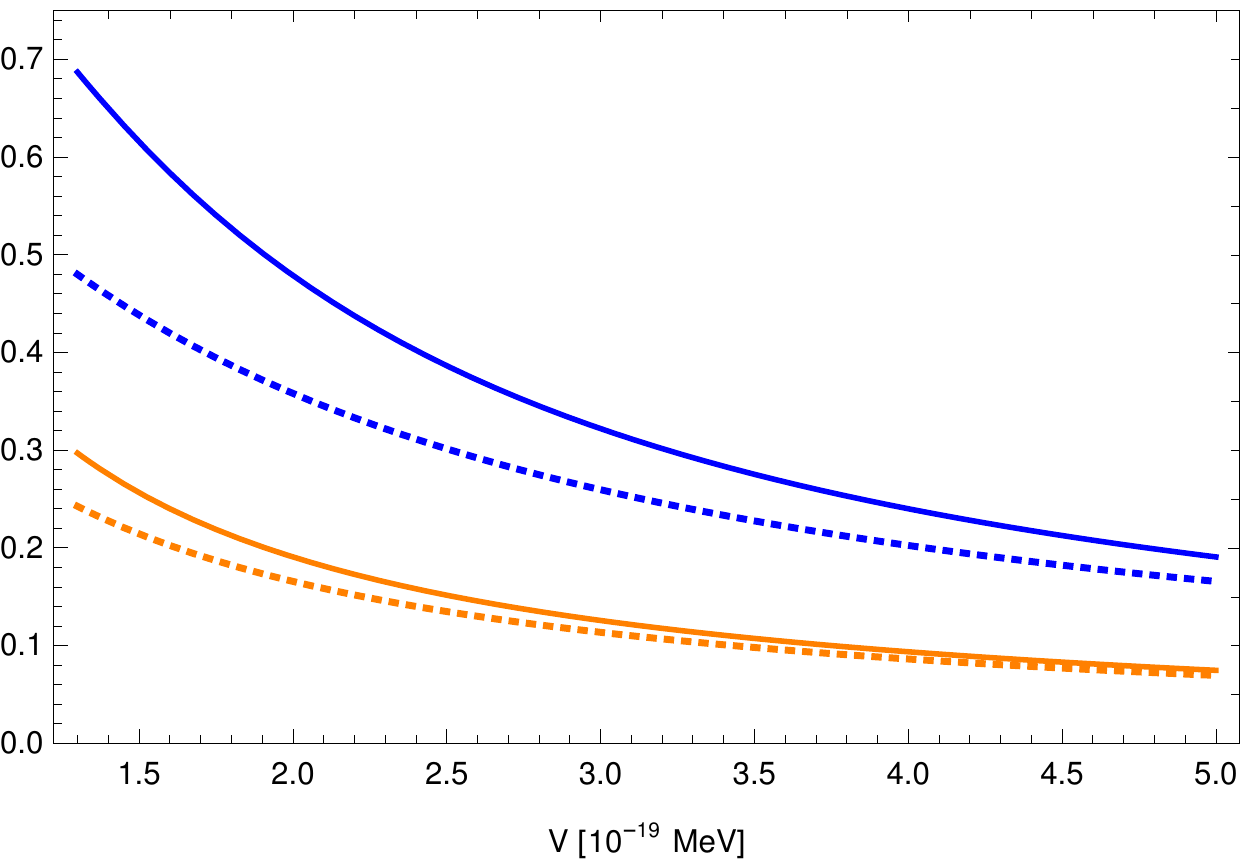}
\end{center}

\caption{$\sin 2 \theta_{12}^m$ in the matter for neutrinos (solid
  lines) and antineutrinos (dashed lines) as the function of their
  interaction potential with Earth for energy $E=400$ MeV (blue lines)
  and $E=1000$ MeV (orange lines).  Plots for the normal and inverted
  hierarchy do not differ.
 \label{fig:angleveff}}

\end{figure}

In the next step, one can obtain analytical formulae for the angles
$\phi_X,\alpha_X$.  We observe that (using the formulae from
ref.~\cite{Ioannisian:2018qwl}) in the limit of large $E\,V$ product
the quantity $\sin 2\theta_{12}^m$ can be expanded as
\bea
\sin 2\theta_{12}^m &=& \frac{ \cos\theta_{13} \sin 2 \theta_{12}}{2
  \cos^2 2 \theta_{13}} \, \frac{\Delta m^2_\odot}{E V} + {\cal
  O}\left( \left(\frac{\Delta m^2_\odot}{E V}\right)^2
\right)
\label{eq:t12exp}
\eea
Therefore, for increasing energy $\sin 2\theta_{12}^{m}$ is suppressed
approximately by $1/(E\, V)$ factor (as illustrated in
Fig.~\ref{fig:angleveff}).  For sufficiently high values of $E$ one
can expand the product~(\ref{eq:xfull}) using the expression for $X_i$
matrices given by eq.~(\ref{eq:xnudef}) and keeping at most the terms
linear in $\epsilon_i \equiv \sin 2\theta_{12}^{mi}$.  Direct
multiplication leads then to the remarkably compact formulae
\bea
\phi_X &=& \nu_1 + \nu_2 + \ldots + \frac{1}{2}\nu_k \nonumber\\
\sin\alpha_X &=&  (\epsilon_k-\epsilon_{k-1})\sin\frac{\nu_k}{2} +
(\epsilon_{k-1}-\epsilon_{k-2})\sin\left(\nu_{k-1} + \frac{\nu_k}{2}
\right) + \ldots \nonumber\\
&+& (\epsilon_2-\epsilon_1)\sin\left(\nu_2 + \nu_3 + \ldots +
\frac{\nu_k}{2} \right) + \epsilon_1 \sin\left(\nu_1 + \nu_2 + \ldots
\frac{\nu_k}{2} \right)
\label{eq:phialexp}
\eea
where the quantities $\nu_i$ and $\epsilon_i$ can be calculated by
numerical diagonalization of the neutrino mixing matrices in Earth
layers or, to a very good accuracy, using the approximate formulae of
ref.~\cite{Ioannisian:2018qwl}.

Such an approximation works well even for energy as low as 300 MeV and
large values of $\sin 2\theta_{12}^{mi}$.  This can be attributed to
the fact that the neglected higher order terms in
eq.~(\ref{eq:phialexp}) are suppressed by additional $\epsilon_i^2$
factors. Eq.~(\ref{eq:phialexp}) reproduces correctly the values of
$\alpha_X$ and $\phi_X$ derived from the numerical calculation of the
matrix $A$ (see Appendix~\ref{app:numphi}) up to about
$E=2$~GeV. However, as we have stressed earlier, for energies $E\gsim
1$ GeV the probabilities defined in eq.~(\ref{eq:PAB}) do not agree
well with the formulae~(\ref{eq:pnonaver}) and the parametrization in
terms of $\alpha_X, \phi_X$ is not useful any more.

Eq.~(\ref{eq:phialexp}) has some remarkable properties.  Firstly, it
shows that the overall neutrino oscillation phase is just a direct sum
of phases in all layers.  In addition, one can check that for neutrino
energies in the range $(300-1000)$ MeV the difference between the
eigenvalues of the Hamiltonian defined in eq.~(\ref{eq:electroweak})
becomes almost constant:
\bea
{\cal H}_2 - {\cal H}_1 \approx V \cos^2  \theta_{13}
%
\label{eq:h12limit}
\eea
For $E>300$ MeV energy-dependent corrections to
eq.~(\ref{eq:h12limit}) are small and the phases $\nu_i$ in
eq.~(\ref{eq:phialexp}) (thus also the overall phase $\phi_X$) depend
to a good approximation only on the azimuthal angle and the Earth
layers density:
\bea
\nu_i \approx V_i \cos^2 \theta_{13} \Delta x_i(\theta)
\label{eq:nuconst}
\eea
where the explicit formulae for the oscillation lengths $\Delta
x_i(\theta)$ are given in Appendix~\ref{app:trig}.

\begin{figure}[htb!]
\begin{center}
\begin{tabular}{cc}
\includegraphics[width=0.44\textwidth]{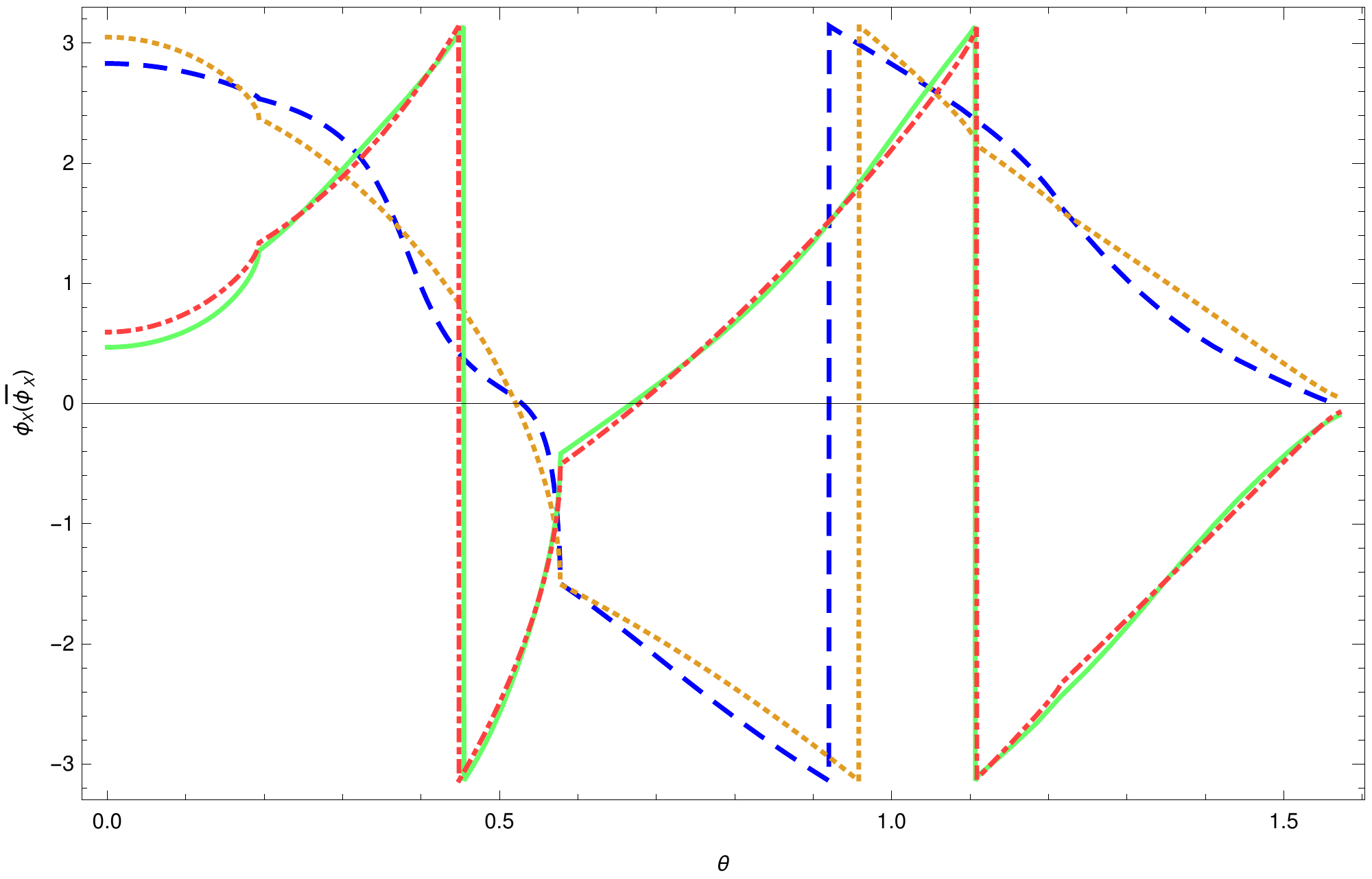} &
\includegraphics[width=0.44\textwidth]{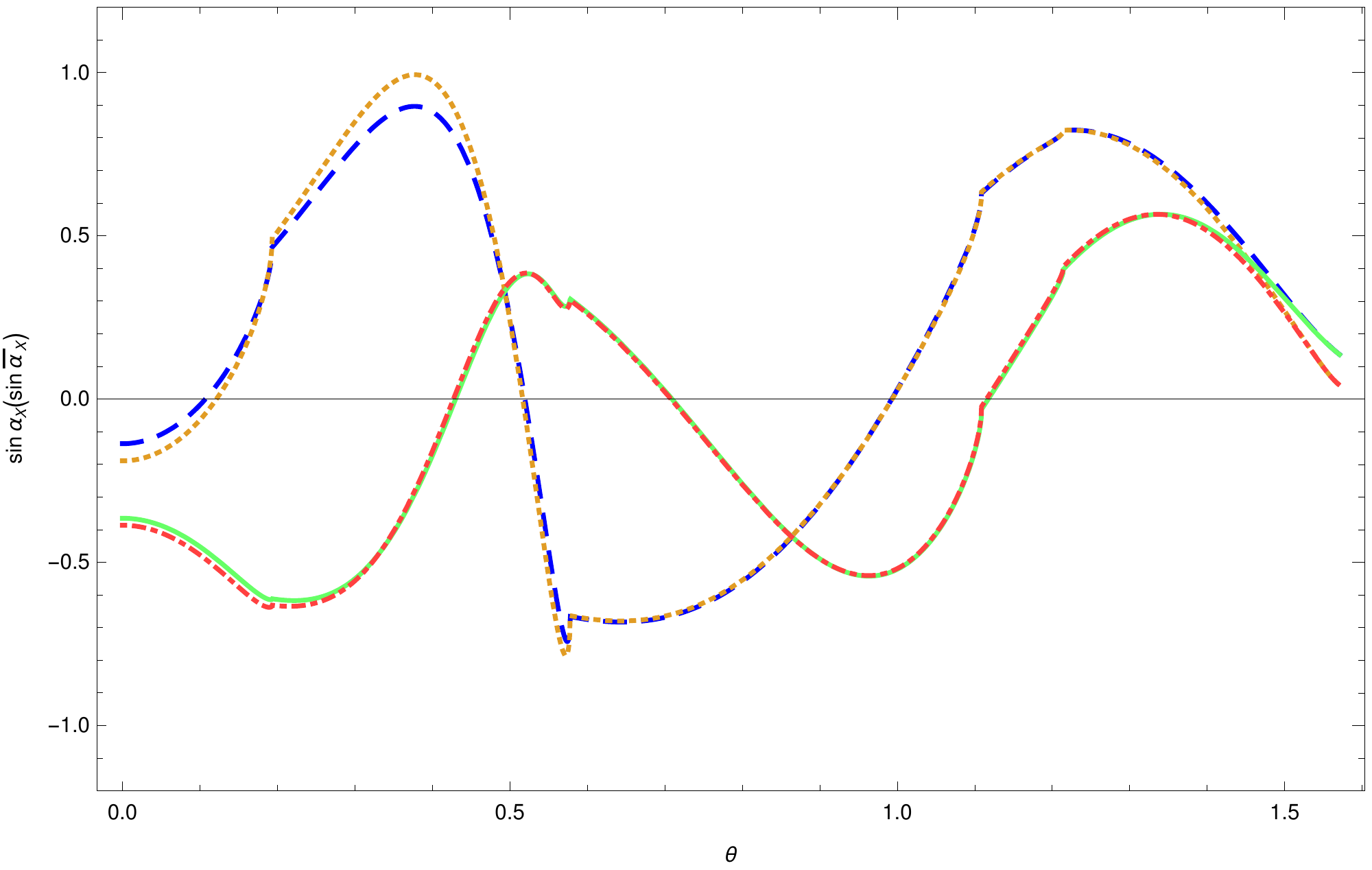}
\\ \includegraphics[width=0.44\textwidth]{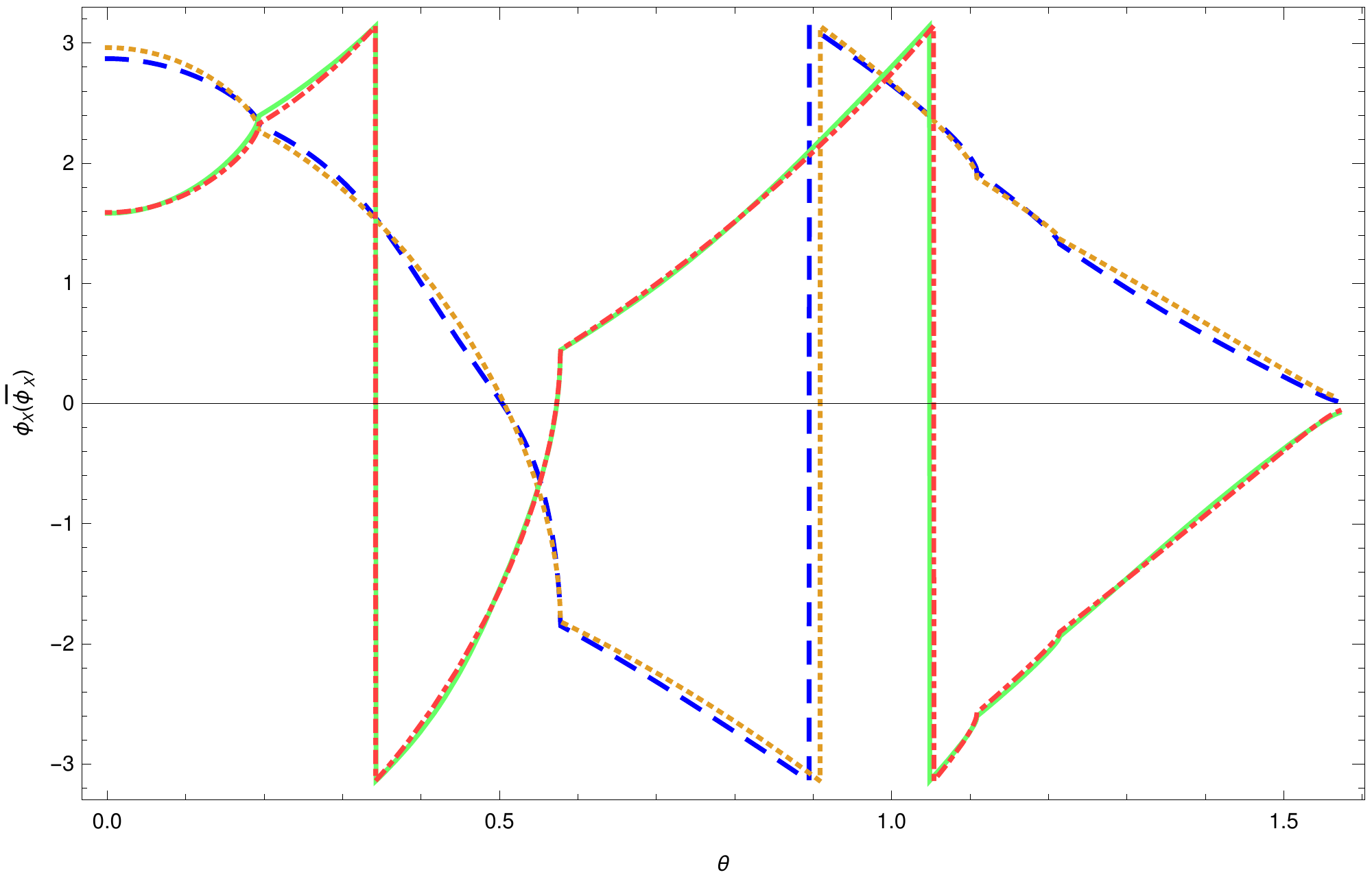} &
\includegraphics[width=0.44\textwidth]{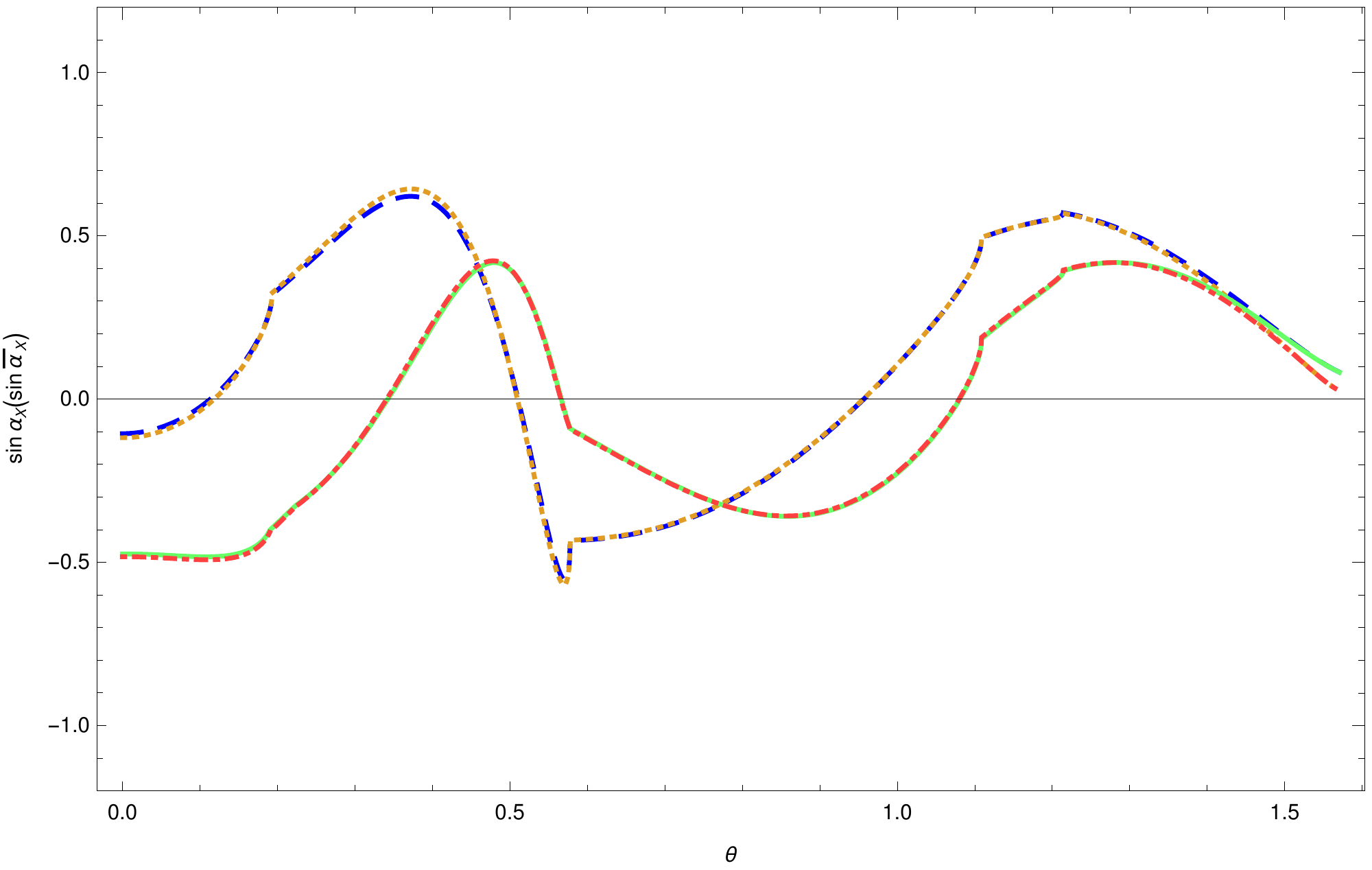} \\
\end{tabular}
\end{center}
\caption{The left panels show comparison of $\phi_X$ obtained from
  numerical diagonalization (blue dashed lines) vs. the approximate
  formulae of eq.~(\ref{eq:phialexp}) (orange dotted lines) and
  similarly for $ \bar\phi_X$ and eq.~(\ref{eq:antiphialexp}) (green
  solid and red dot-dashed line, respectively) as a function of the
  azimuthal angle.  The right panels show the analogous comparison of
  $\sin\alpha_X$ and $\sin\bar\alpha_X$.  Chosen neutrino energies are
  $E = 300$ MeV for the upper row and $E= 500$ MeV for the lower
  row. Normal mass ordering is assumed.
  \label{fig:phialexp}}
\end{figure}

\begin{figure}[htb!]
\begin{center}
\begin{tabular}{cc}
\includegraphics[width=0.44\textwidth]{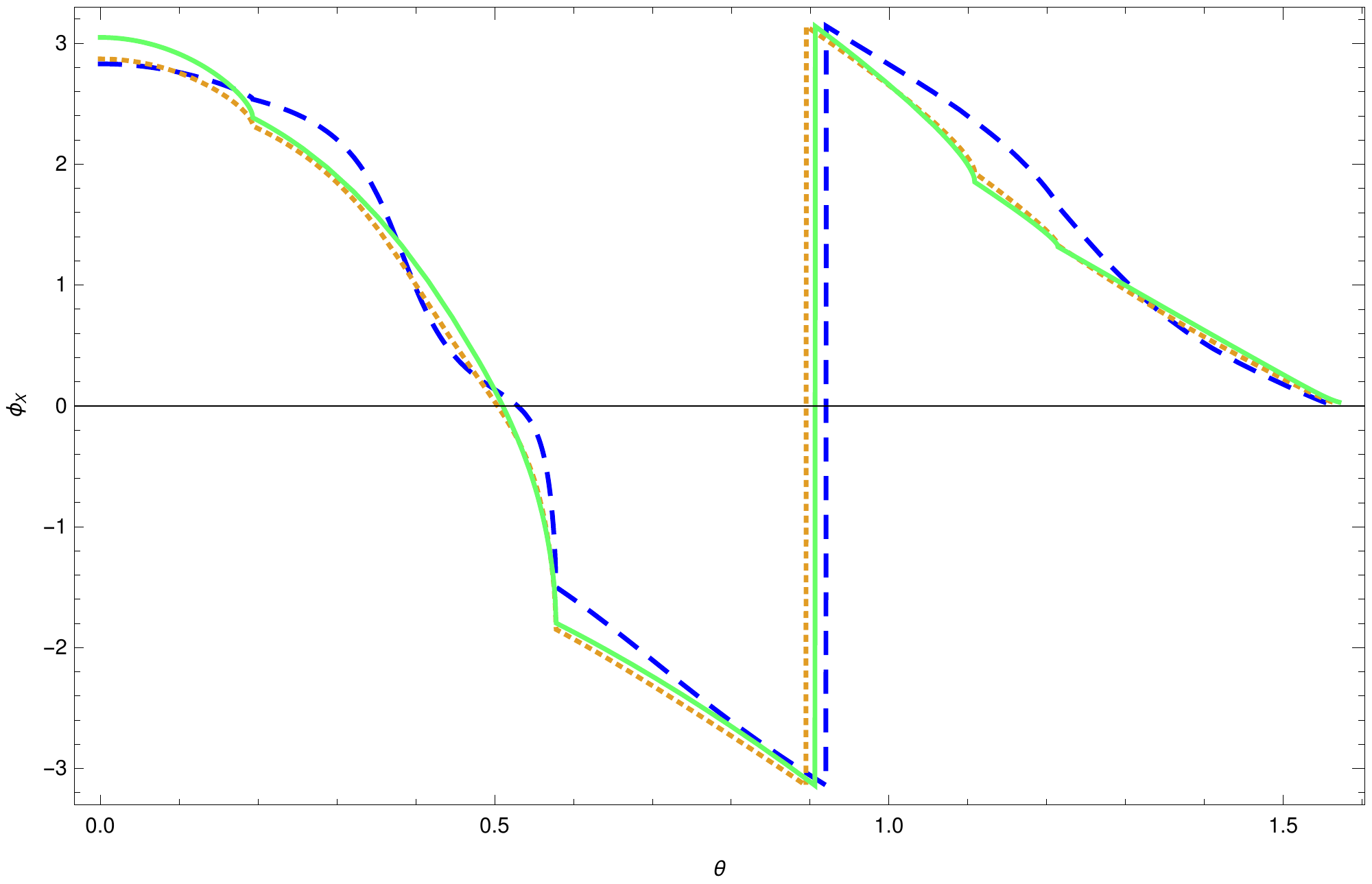} &
\includegraphics[width=0.44\textwidth]{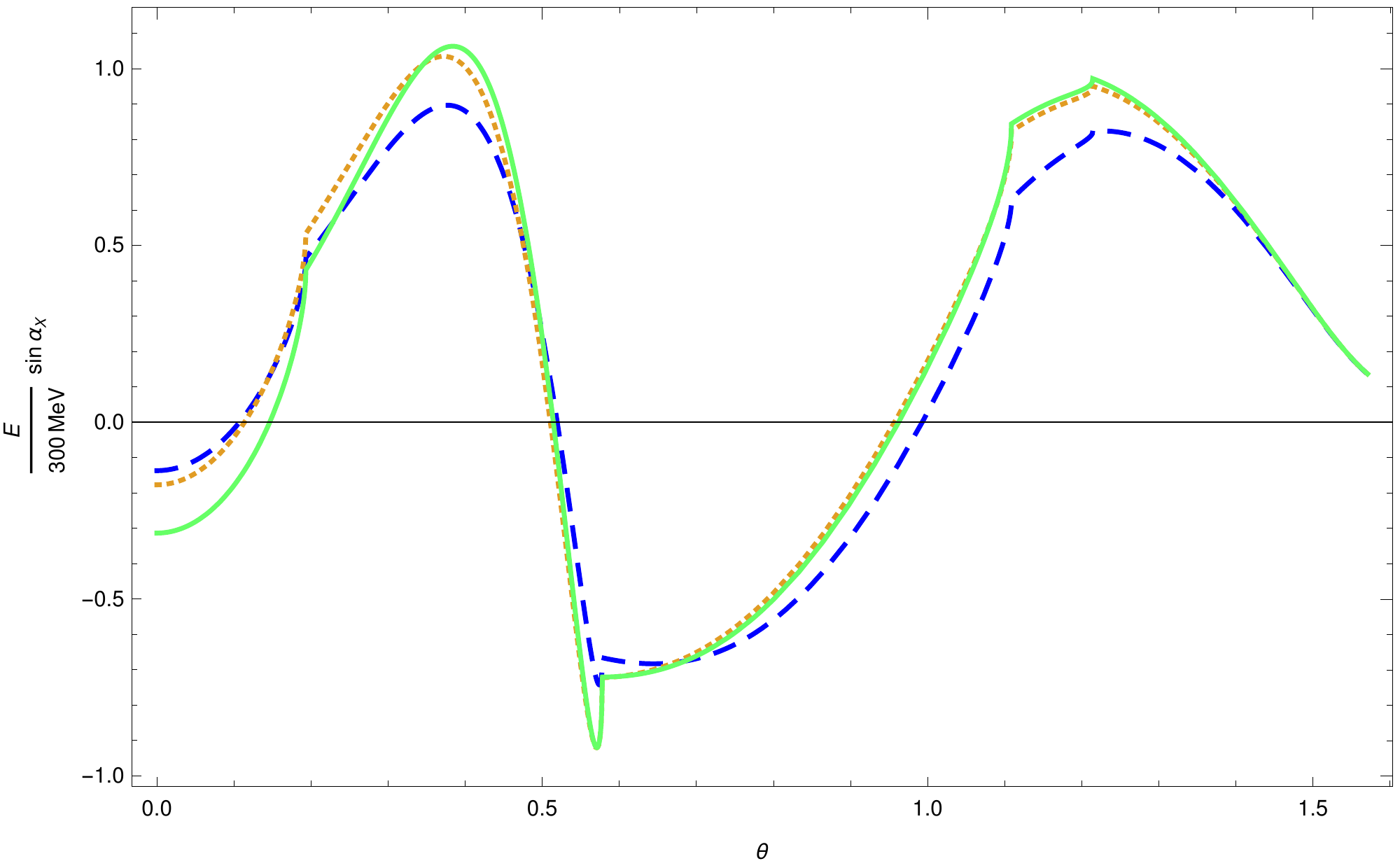}\\
\includegraphics[width=0.44\textwidth]{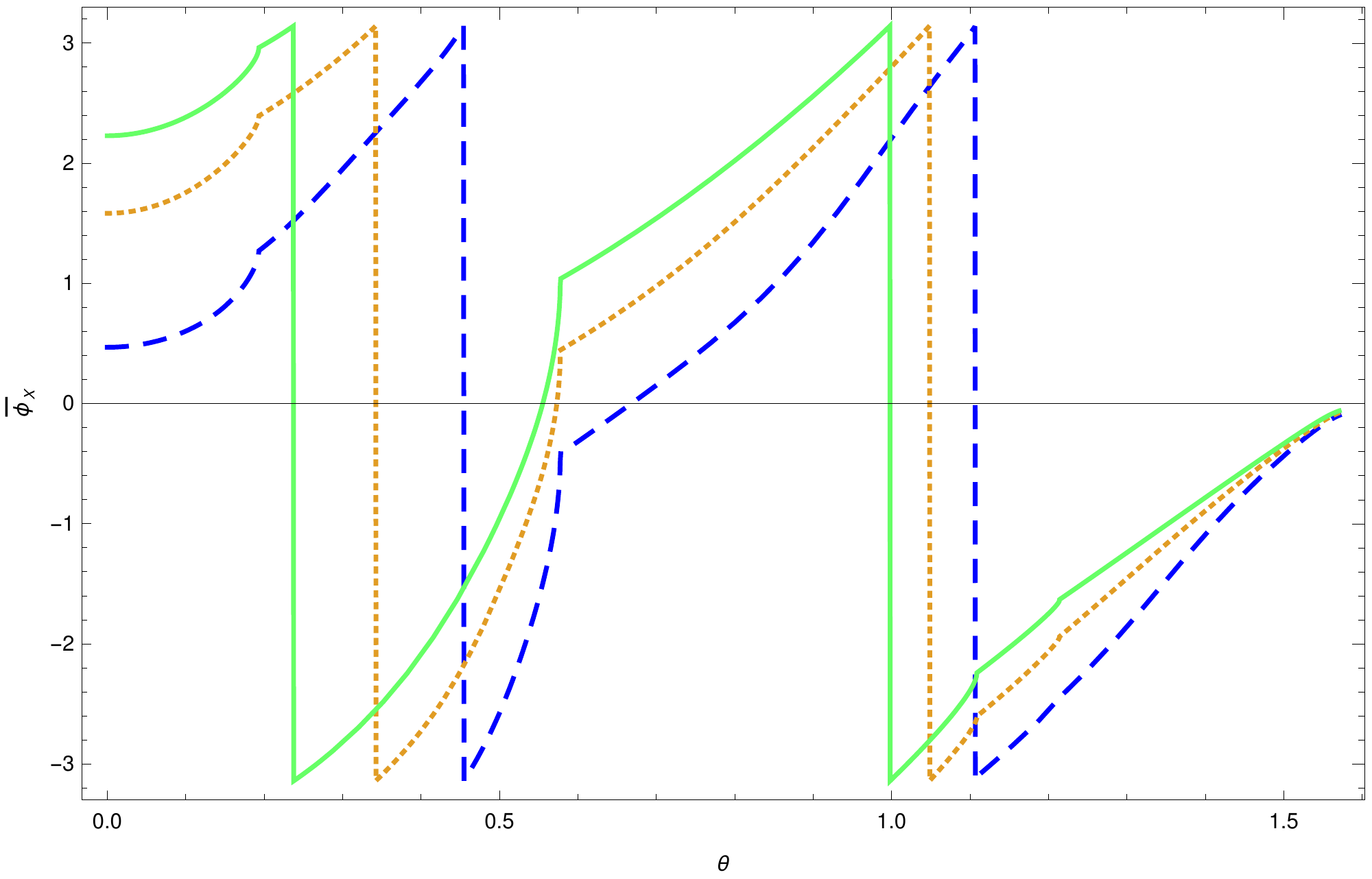} &
\includegraphics[width=0.44\textwidth]{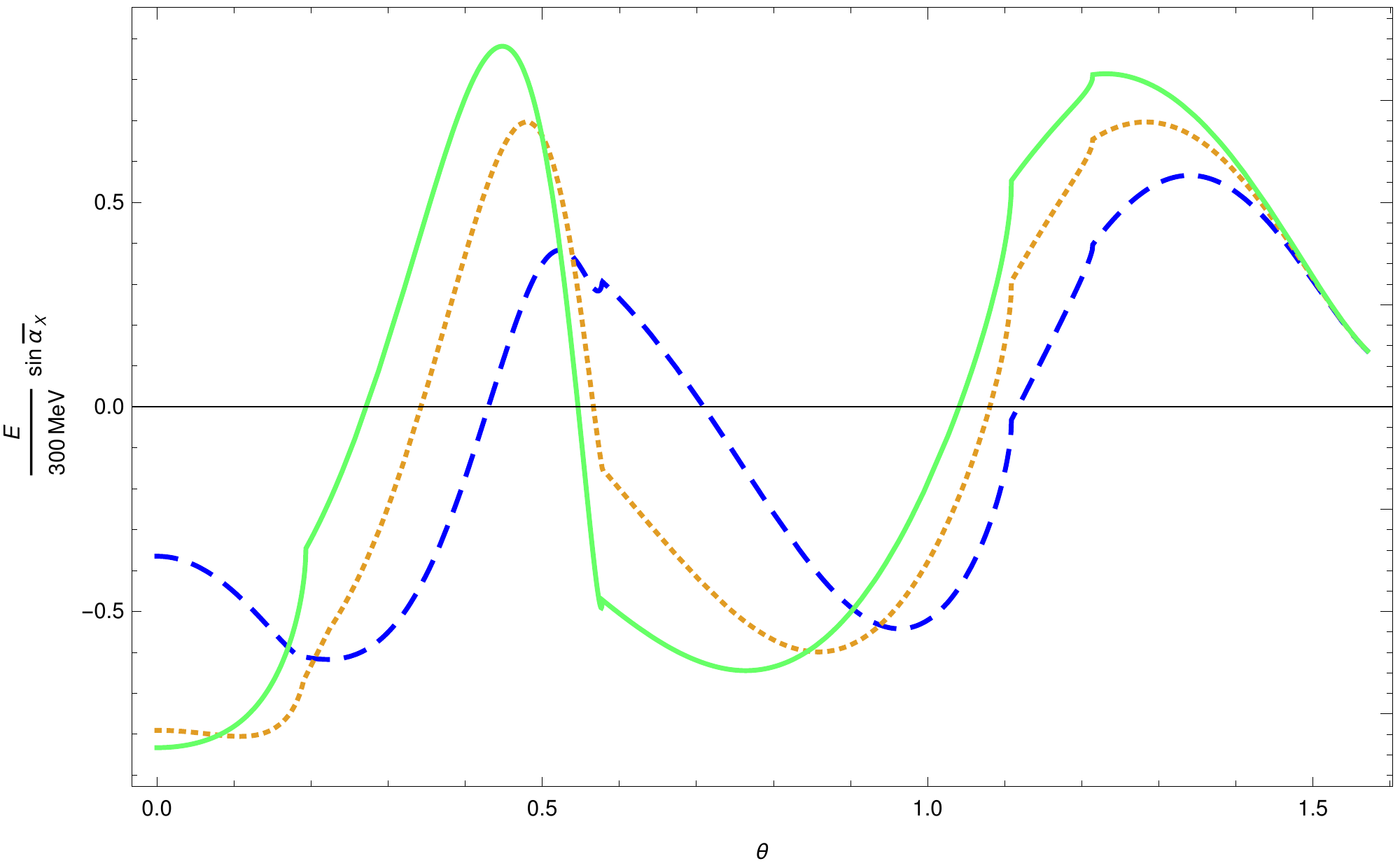} \\
\end{tabular}
\end{center}
\caption{Dependence of $\phi_X, \bar\phi_X, \frac{E}{300~\mathrm{MeV}}
  \sin\alpha_X$ and $\frac{E}{300~\mathrm{MeV}} \sin\bar\alpha_X$
  obtained from numerical diagonalization on the azimuthal angle for
  the neutrino energies $E=300, 500$ and $1000$ MeV (blue dashed,
  orange dotted and green solid line, respectively).  Normal mass
  ordering is assumed.
  \label{fig:phialscale}}
\end{figure}

Secondly, since $\epsilon_i \equiv \sin 2\theta_{12}^{mi} \sim 1/E$
and the phases $\nu_i$ become energy independent, for $E>300$ MeV to a
good approximation one has $\sin\alpha_X = f(\theta)/E$, where
$f(\theta)$ is some function of the azimuthal angle only.

Similar approximation holds for the antineutrino oscillations, for
which one needs to replace $V\to - V$ and $\delta\to - \delta$ in
eq.~(\ref{eq:electroweak}). In this case effective $\bar\phi_X$ and
$\sin\bar\alpha_X$ parameters (we denote all variables related to
antineutrino oscillations with barred symbols) read as:
\bea
\bar\phi_X &=& \bar\nu_1 + \bar\nu_2 + \ldots + \frac{1}{2}\bar\nu_k
\nonumber\\
\sin\bar\alpha_X &=&
- (\bar\epsilon_k-\bar\epsilon_{k-1})\sin\frac{\bar\nu_k}{2} -
(\bar\epsilon_{k-1}-\bar\epsilon_{k-2})\sin\left(\bar\nu_{k-1} -
\frac{\bar\nu_k}{2} \right) + \ldots \nonumber\\
&-& (\bar\epsilon_2-\bar\epsilon_1)\sin\left(\bar\nu_2 + \bar\nu_3 +
\ldots + \frac{\bar\nu_k}{2} \right) - \epsilon_1 \sin\left(\bar\nu_1
+ \bar\nu_2 + \ldots + \frac{\bar\nu_k}{2} \right)
\label{eq:antiphialexp}
\eea
with
\bea
\bar\nu_i &=& (\bar{\cal H}^i_2 - \bar{\cal H}^i_1)\Delta
x_i
\nnb\\[2mm]
\bar\epsilon_i &=& \sin 2 \bar\theta_{12}^{mi}
\label{eq:antinudef}
\eea
The accuracy of approximation~(\ref{eq:antiphialexp}) is even better
than that of eq.~(\ref{eq:phialexp}) due to the hierarchy of the
expansion parameters, $\bar\epsilon_i<\epsilon_i$ (see
Fig.~\ref{fig:angleveff}).

The dependence of $\phi_X, \bar\phi_X$ and $\sin \alpha_X,
\sin\bar\alpha_X$ on energy and azimuthal angle and the comparison of
numerical fitting (see Appendix~\ref{app:numphi}) and analytical
approximate formulae of eqs.~(\ref{eq:phialexp},
\ref{eq:antiphialexp}) for these parameters is illustrated in
Fig.~\ref{fig:phialexp} (where the normal neutrino mass ordering is
assumed).  As can be seen, for $E>300-400$ MeV numerical and
analytical results agree very well.

Fig.~\ref{fig:phialscale} shows how the effective parameters are
modified when neutrino or antineutrino energy changes. As discussed
above, the dependence of the $\phi_X$ and $\sin\alpha_X$ on the angle
$\theta$ becomes universal with energy, up to an overall $1/E$ scaling
of the $\sin\alpha_X$ amplitude.  For $\bar\phi_X$, there remains much
stronger energy dependence and the $1/E$ scaling of $\sin\bar\alpha_X$
is less exact. This is a consequence of the fact that in the sub-GeV
range the energy-dependent corrections to the approximation $\bar{\cal
  H}^i_2 - \bar{\cal H}^i_1\approx -V_i\cos^2\theta_{13}$ are
significantly larger than for the neutrino case.

The dependence of $\phi_X,\bar\phi_X$ and $\sin \alpha_X, \sin
\bar\alpha_X$ on the azimuthal angle for the inverse mass ordering is
almost identical, with small differences of the order of few \%
appearing only for small values of $\theta$.

\subsection{Energy and the angular dependence of
  the oscillation probabilities}
\label{sec:phdep}

As already mentioned in the previous Section, the dependence of
$\phi_X$ and $E \, \times \, \sin\alpha_X$ (and similarly for
antineutrinos) on the azimuthal angle is almost identical for the
normal and inverted neutrino mass ordering, thus the problem of the
CP-phase determination is not affected by the assumption of the mass
hierarchy~\cite{Indumathi:2017kxa}.

Using the parametrization of eq.~(\ref{eq:asolve}), for energies
larger than $300-400$ MeV one then obtains very compact expressions
for the averaged oscillation probabilities.  Experimentally detected
numbers of electron and muon neutrinos (antineutrinos) are
proportional (taking into account their different atmospheric fluxes
$N_{\nu_\mu} \approx 2 N_{\nu_e}$) to quantities defined as
\bea
I_e &=& I_{ee} + 2\; I_{\mu e} \nn
I_\mu &=& I_{\mu \mu} + \frac{1}{2}\; I_{e \mu}
\label{eq:pepmu}
\eea

From eqs.~(\ref{eq:asolve}) we see that $I_{ee}$ does not depend on
the CP-phase.  That fact and, in addition, the difference in the
electron and muon neutrino (antineutrino) fluxes, make the observable
$I_e$ much more efficient than $I_\mu$ for measuring the CP phase
$\delta$ (and similarly for antineutrinos). The latter quantity has
some additional (even) CP phase dependence in $I_{\mu\mu}$ that masks
the (odd) dependence on the CP phase of $I_{e \mu}$.  Assuming the
central values for the measured vacuum mixing angles (and the normal
mass ordering), for the transition probabilities we get the simple but
very accurate approximation:
\bea
I_e &\approx& 1.00 - 0.094 \sin^2\alpha_X - 0.143 \sin2 \alpha_X
\sin(\delta + \phi_X)\nnb\\[2mm]
I_\mu &\approx& 0.50 + 0.011 \cos 2 \phi_X + 0.011 (1 - \cos 2 \phi_X)
\sin^2\alpha_X + 0.021 \sin^2\alpha_X \cos^2\delta \nnb\\
&+& (0.036 \cos \phi_X \sin\delta +  0.029 \sin \phi_X \cos\delta ) \sin
2 \alpha_X
\label{eq:pmesolve1}
\eea
The same equations are valid for antineutrinos, after replacing
$\delta\to -\delta$ and $\phi_X,\alpha_X\to \bar\phi_X,\bar\alpha_X$.

As an immediate consequence of eq.~(\ref{eq:pmesolve1}) we observe
that the variation of $\bar P_e$ with the CP phase $\delta$ cannot be
larger than $ \approx 0.30$.  Furthermore, the effects of CP
violation, which as can be seen, are proportional to $\sin 2\alpha_X$,
decrease approximately like $1/E$ (as it is discussed earlier, this
scaling is less exact for antineutrinos). The CP phase dependence of
$I_\mu$ is much weaker. Both are shown in Fig.~\ref{fig:aver} for some
optimal values of the azimuthal angle, to be discussed in the next
section.
 
For any given energy $E$ and the azimuthal angle $\theta$ the
parameters $\phi_X,\bar\phi_X,\sin\alpha_X$ and $\sin\bar\alpha_X$ are
calculable using either the numerical diagonalization of neutrino
Hamiltonian and fitting procedure described in
Appendix~\ref{app:numphi} or, for sufficiently large $E$, the
approximate formulae of eq.~(\ref{eq:phialexp}) and they depend only
on the assumed Earth density profile.  Therefore, as follows from
eq.~(\ref{eq:pmesolve1}), one can subtract the theoretically known
CP-independent terms from the experimentally measured transition
probabilities and obtain directly the constraints on the combination
$\delta + \phi_X(E,\theta)$ or on the products $\cos \phi_X
\sin\delta, \sin \phi_X \cos\delta$.  Performing a fit to many bins in
energy and azimuthal angle one can determine the value of phase
$\delta$ itself.

\section{Optimal observables for the CP-phase detection}
\label{sec:genopti}

\subsection{Optimal azimuthal angles}
\label{sec:angopti}

\begin{figure}[tb!]
\begin{center}
\begin{tabular}{cc}
\includegraphics[width=0.45\textwidth]{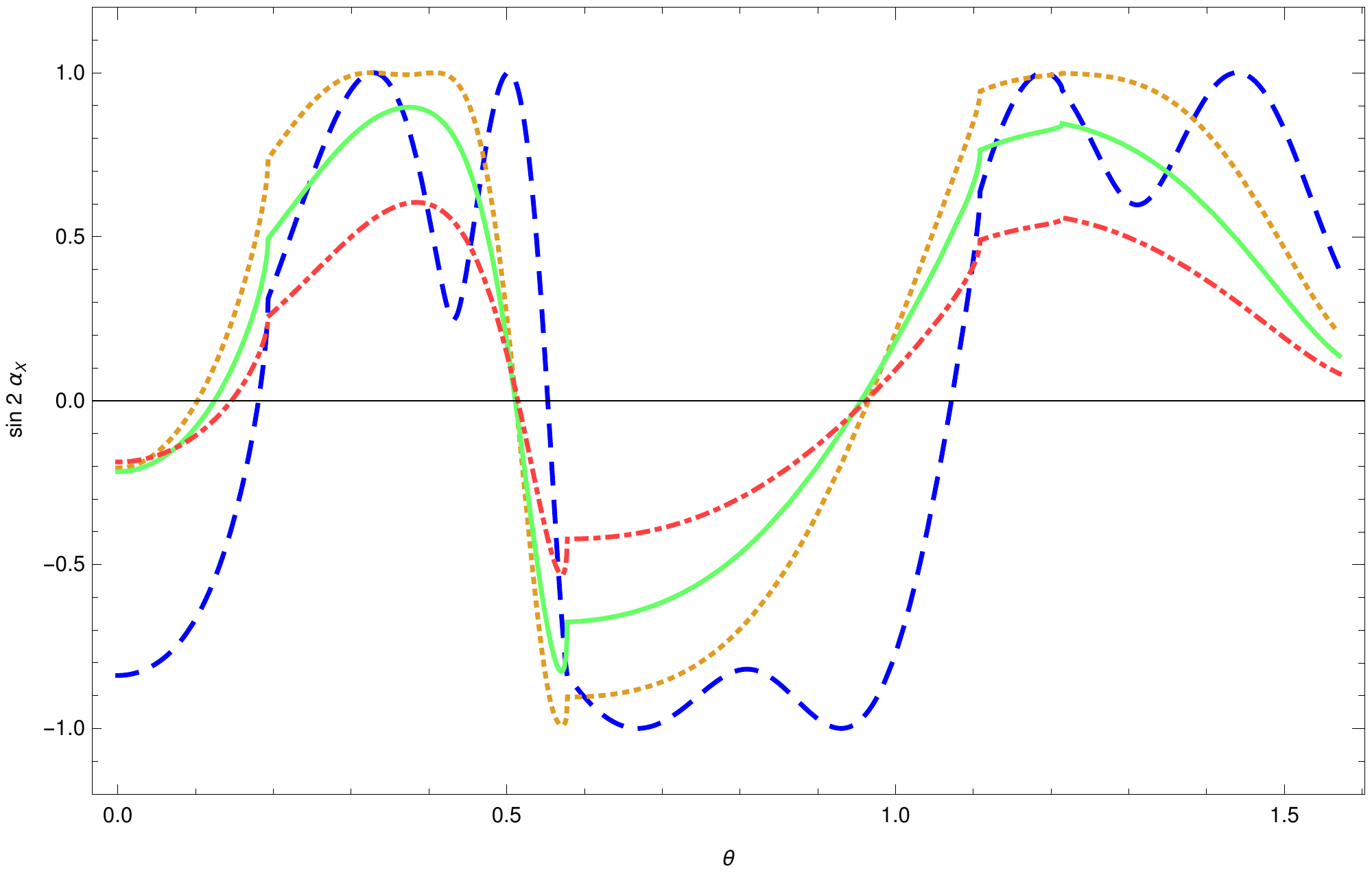} &
\includegraphics[width=0.45\textwidth]{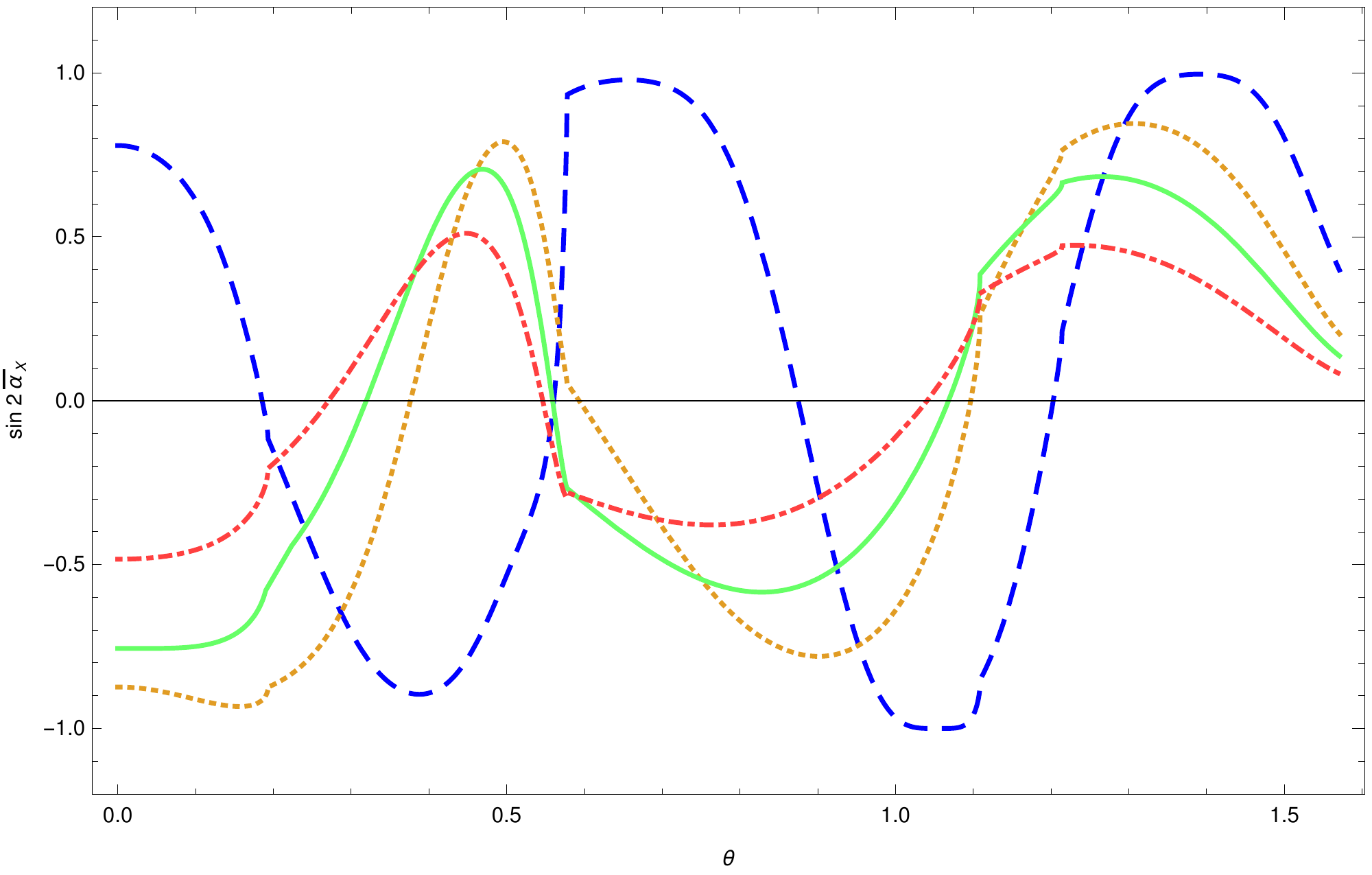}
\end{tabular}
\end{center}
\caption{ $\sin 2 \alpha_X$ and $\sin 2 \bar \alpha_X$ plotted as a
  function of the azimuthal angle varied from 0 to $\pi/2$ for
  neutrino energies $E=200$ MeV (blue dashed line), $E=400$ MeV
  (orange dotted line), $E=600$ MeV (green solid line) and $E=1000$
  MeV (red dot-dashed line).  Normal mass ordering is
  assumed.  \label{fig:sin2alpha}}

\end{figure}

The experimental chances of measuring the CP-phase in
$\nu_\mu\to\nu_e(\bar\nu_\mu\to\bar\nu_e)$ transitions are best when
the coefficients of CP-violating terms are maximal.  This happens when
the $\sin 2 \alpha_X(\sin 2\bar \alpha_X$) reaches maximal or minimal
value.  In Fig.~\ref{fig:sin2alpha} we plot the dependence of $\sin 2
\alpha_X(E,\theta)(\sin 2\bar \alpha_X$) as a function of the
azimuthal angle for few chosen values of neutrino energy.  As can be
seen, independently of the neutrino energy, the extreme values of
$\sin 2 \alpha_X(E,\theta)$ are reached for three values of the
azimuthal angle $\theta_1 = 0.12\pi, \theta_2 = 0.18\pi,
\theta_3=0.39\pi$ and for these three values variations $I_{ e},
I_\mu$ with the phase $\delta$ are maximal and give the best chance
for its successful measurement.  Extreme values of $\sin 2\bar
\alpha_X$ are more energy dependent, however they can be easily
calculated, numerically or analytically from
eq.~(\ref{eq:antiphialexp}), for any chosen energy value. Therefore,
although in what follows we concentrate on discussing the neutrino
oscillations, the results can be in a straightforward way extended to
the case of antineutrino mixing, leading to similar conclusions.

\begin{figure}[tb!]
\begin{center}
\begin{tabular}{cc}
\includegraphics[width=0.45\textwidth]{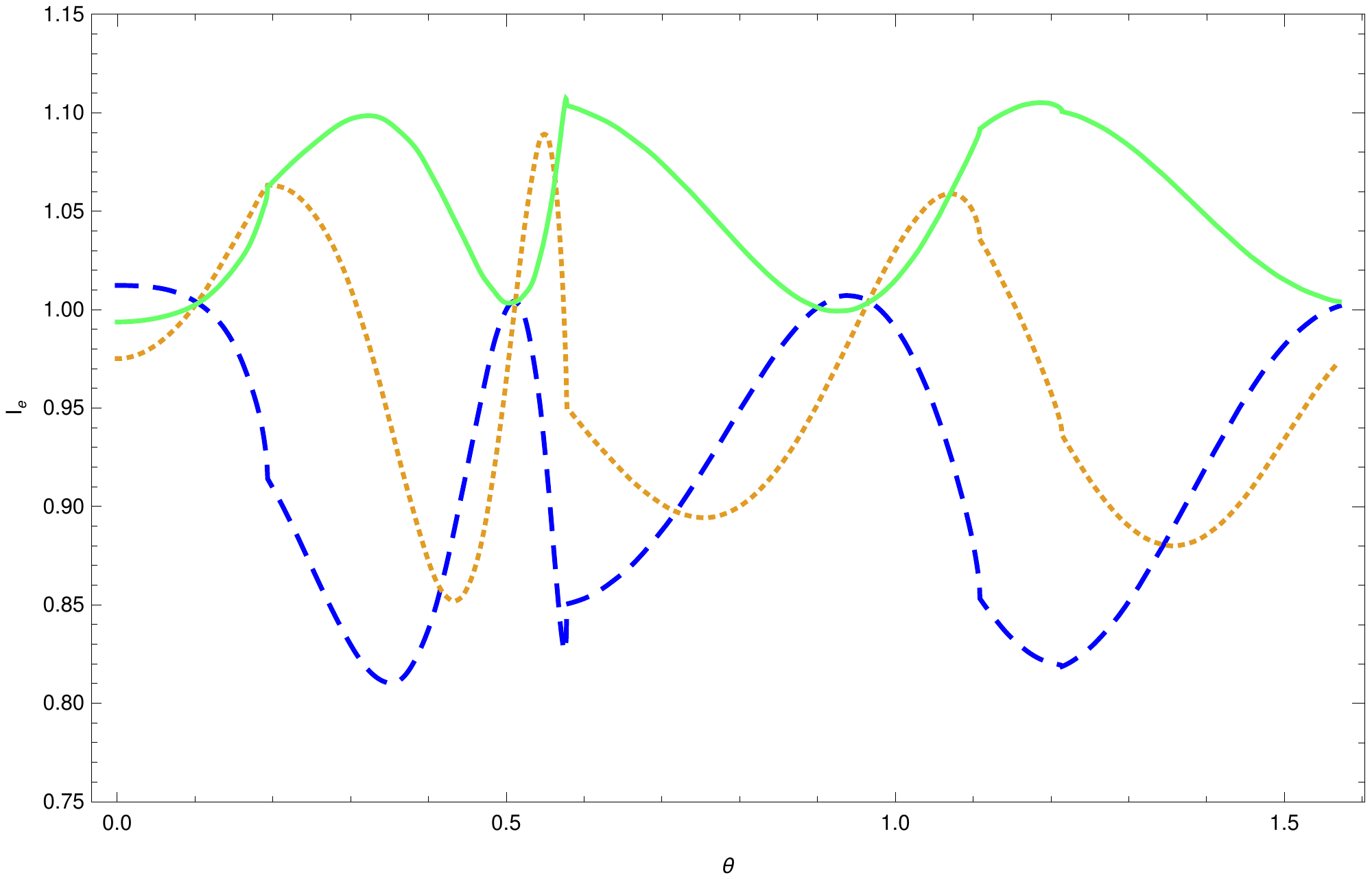} &
\includegraphics[width=0.45\textwidth]{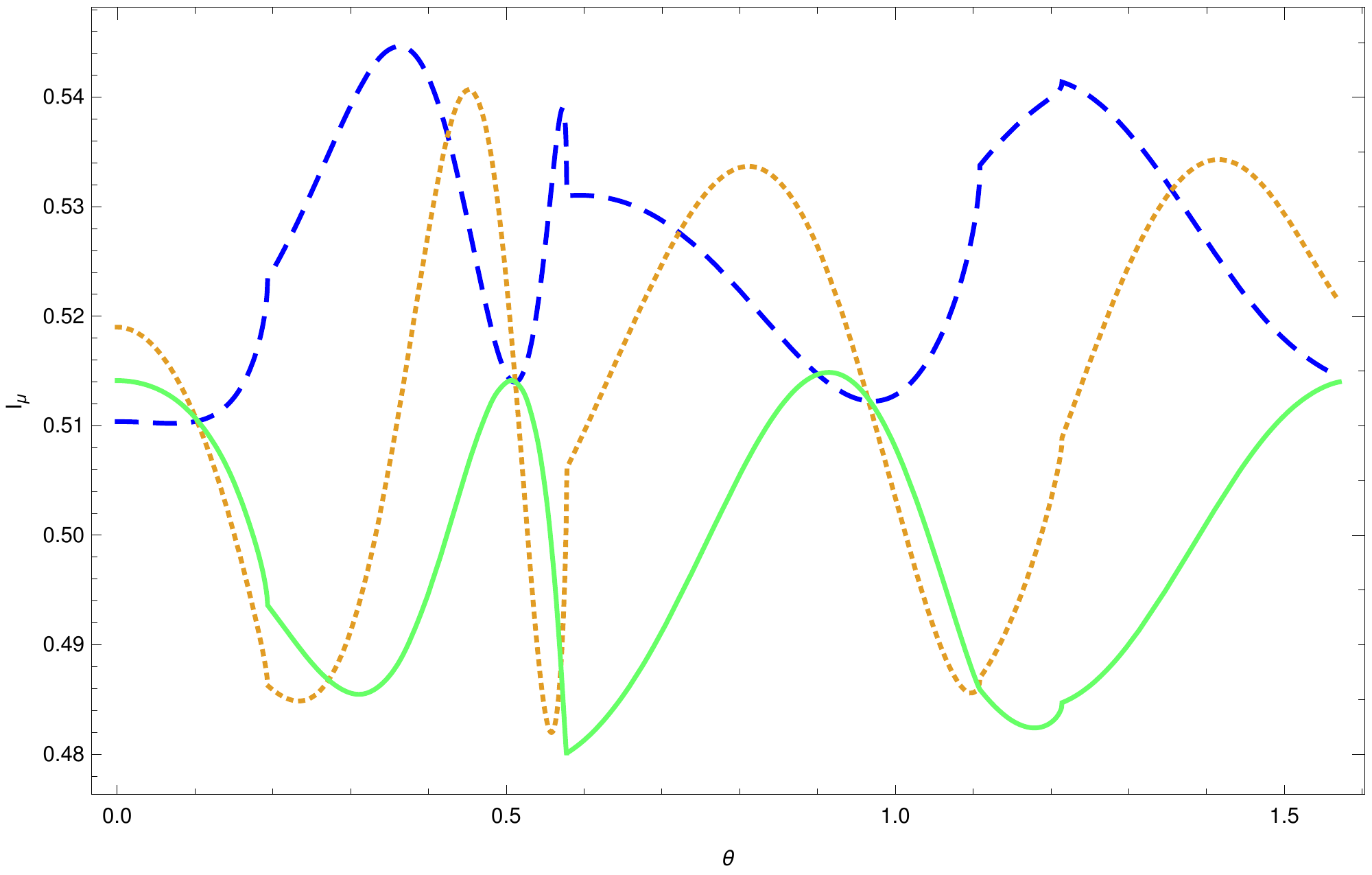}
\end{tabular}
\end{center}
\caption{$I_{e}$ (left panel) and $I_\mu$ (right panel) for neutrinos
  as a function of the azimuthal angle varied from 0 to $\pi/2$ and
  fixed $E=400$ MeV. Blue dashed line: $\delta=0$; orange dotted line:
  $\delta=\pi/2$; green solid line: $\delta=\pi$.  Normal mass
  ordering is assumed.  \label{fig:aver}}

\end{figure}

In Fig.~\ref{fig:aver}, where we plot the dependence of the quantities
$I_{ e}, I_\mu$ for neutrinos on the azimuthal angle for $E=400$ MeV
and several values of the phase $\delta$ assuming the normal neutrino
mass ordering (corresponding plot for the inverse neutrino mass
ordering is almost identical). As already mentioned in the previous
section, $I_e$ is more sensitive to CP-phase than $I_\mu$.  We also
see that there are also "the worst" values of the azimuthal angle
where the dependence on the CP phase vanishes.

\begin{figure}[tb!]
\begin{center}
\begin{tabular}{cc}
\includegraphics[width=0.45\textwidth]{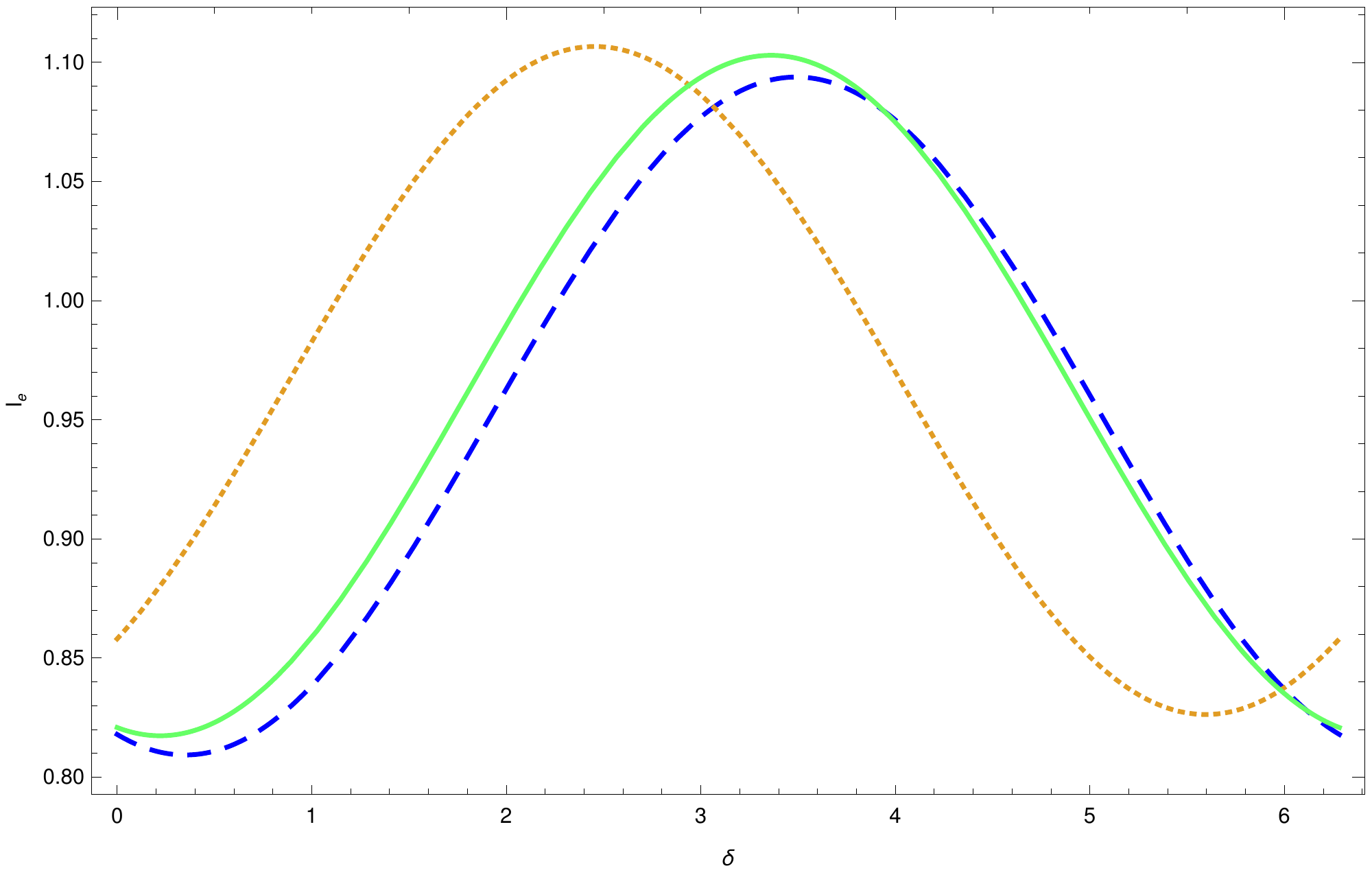} &
\includegraphics[width=0.45\textwidth]{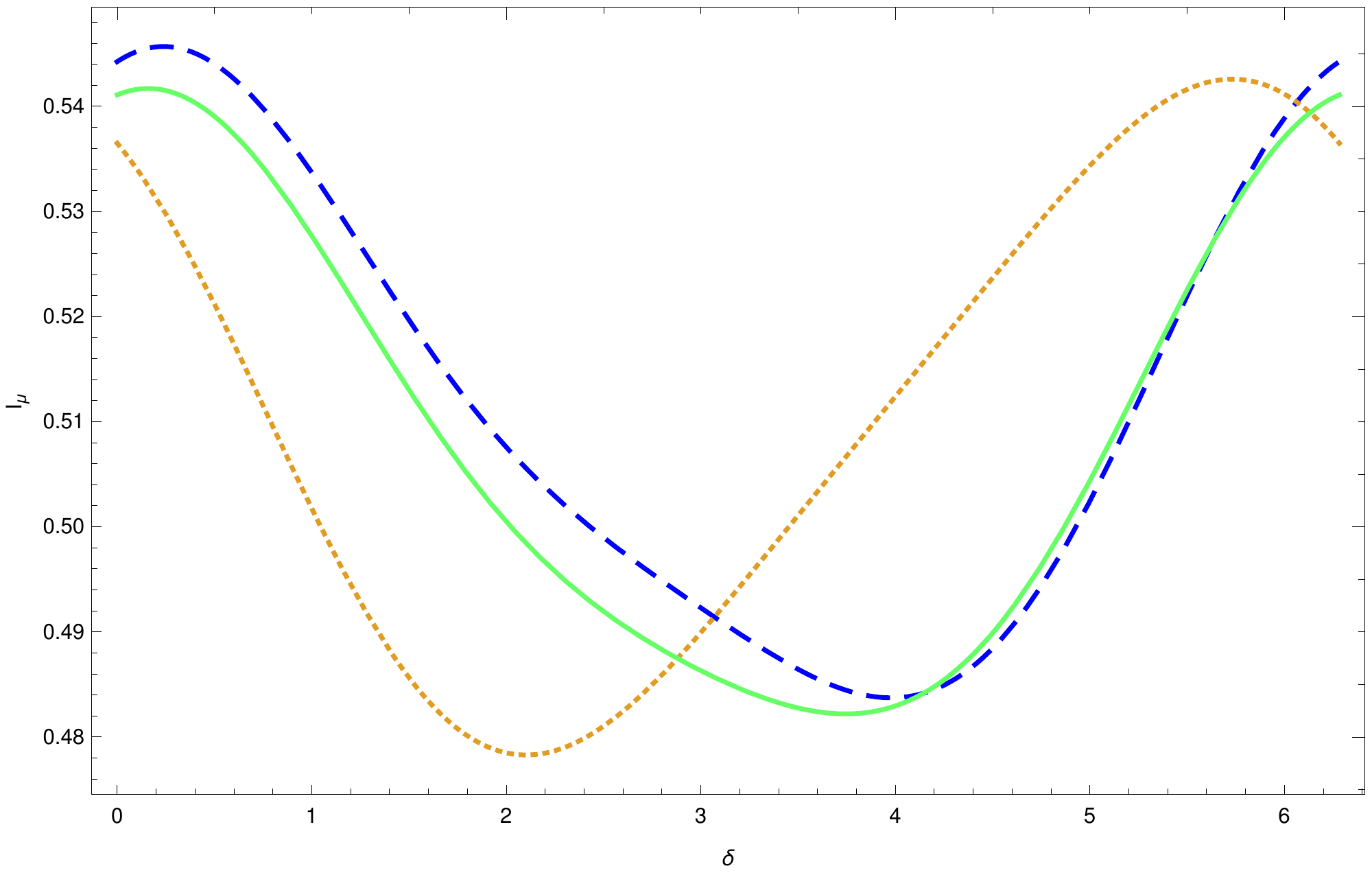}
\end{tabular}
\end{center}

\caption{$I_{e}$ (left panel) and $I_\mu$ (right panel) for neutrinos
  as a function of the CP-phase for $E=400$ MeV and optimal angles
  $\theta_1, \theta_2, \theta_3$ (blue dashed, orange dotted and green
  solid line, respectively).  Normal mass ordering is
  assumed.  \label{fig:enaver}}

\end{figure}

For the angles $\theta_1$ and $\theta_3$, corresponding to maxima of
$\sin 2 \alpha_X$, one has $\phi_X(\theta_1)\approx \phi_X(\theta_3)$
and $\sin\alpha_X(\theta_1)\approx \sin\alpha_X(\theta_3)$.
Therefore, measurements done for $\theta_1$ and $\theta_3$ provide
information on $\delta$.  For the angle $\theta_2$ we get
$\sin\alpha_X(\theta_2)\approx -\sin\alpha_X(\theta_1)$ but different
phase $\phi_X$, thus combining measurements for all three azimuthal
angles gives a chance for determining the phase $\delta$ itself.
Discussed effects are illustrated in Fig.~\ref{fig:enaver}, where we
plot the dependence of $I_e, I_\mu$ for neutrinos as a function of the
CP-phase for $E=400$ MeV and optimal angles $\theta_1, \theta_2,
\theta_3$.  As expected, in this case the variation of $I_e$ due to
the phase dependence reaches maximal allowed value of $0.30$ and the
effect is much weaker for $I_\mu$.  The maxima and minima of the line
corresponding to angle $\theta_2$ are shifted compared to other two
lines due to different value of $\phi_X(\theta_2)$. For antineutrinos
one obtains very similar results, with the optimal values for the
azimuthal angle at $E=400$ MeV being $\theta=0.05\pi, 0.16\pi,
0.29\pi$ and $0.42\pi$.

\subsection{Optimised observables}
\label{sec:opti}

As discussed in the previous Section, eq.~(\ref{eq:asolve}) can be
used to determine the phase $\delta$ by subtracting from the
experimentally measured transition probability the theoretically
calculated CP-independent terms.  The analytical understanding of
$\phi_X$ and $\sin\alpha_X$ behaviour has shown that they have very
simple energy dependence which allows us to design the alternative
observable well suited to measure the CP violating phase, based on
subtracting experimentally measured numbers of $\nu_e$ and $\nu_\mu$,
proportional to quantities $I_e,I_\mu$ defined in eq.~(\ref{eq:pepmu})
and a well know combination of neutrino vacuum mixing angles. In this
case our discussion holds only for the neutrino oscillations, as the
energy scaling of antineutrino mixing probability is more complicated
and does not allow for such a simple cancellations as we are
exploiting below.

Inspection of the expression for $I_e$ which can be derived from
eq.~(\ref{eq:asolve}) shows that it consists of a constant term
depending only on the vacuum oscillation angles, term proportional to
$\sin^2\alpha_X$ but not depending on the phase $\delta$ (which for
$E>300-400$ MeV scales to a very good accuracy like $1/E^2$) and a
term proportional to $\sin(\delta + \phi_X)$ which scales
approximately like $1/E$.  Therefore, for $E_1,E_2>300-400$ MeV and
for {\em any} azimuthal angle $\theta$ the quantity
\bea
\Delta I_e(E_1,E_2,\theta) &=& \frac{E_1^2}{E_2^2} I_e(E_1,\theta) -
I_e(E_2,\theta) - \left(1 - \frac{\sin^2 2 \theta_{13} \cos 2
  \theta_{23}}{2} \right) \left( \frac{E_1^2}{E_2^2} -1 \right) \nnb\\
&\approx& - \cos^2\theta_{13} \sin\theta_{13} \sin 2 \theta_{23}
\left(\frac{E_1^2}{E_2^2}\sin 2 \alpha_X(E_1) - \sin 2
\alpha_X(E_2)\right) \sin(\delta + \phi_X)\nnb\\
&\approx& - 0.14 \left(\frac{E_1^2}{E_2^2}\sin 2 \alpha_X(E_1) - \sin
2 \alpha_X(E_2)\right) \sin(\delta + \phi_X)
\label{eq:deltap21}
\eea
is to a good approximation proportional solely to the sine of the
CP-violating phase shifted by $\phi_X$.  To maximise $\Delta
I_e(E_1,E_2,\theta)$, one can choose $\theta$ equal or close to the
values maximising $|\sin 2 \alpha_X|$, as described in the previous
Section, and large splitting between $E_1$ and $E_2$, like e.g.  $E_1
= 400$ MeV, $E_2=1000$ MeV.

\begin{figure}[tb!]
\begin{center}
\includegraphics[width=0.7\textwidth]{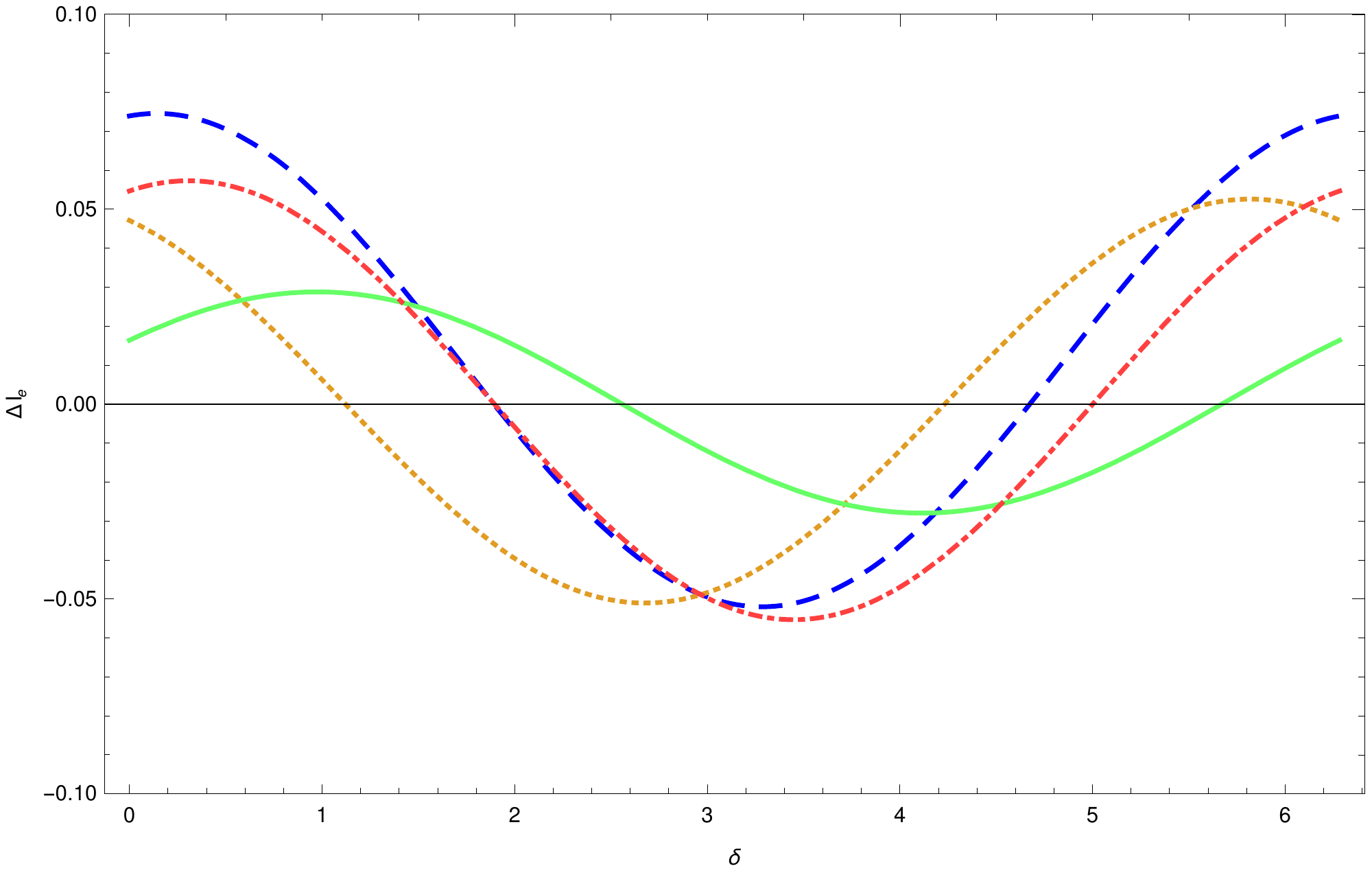}
\end{center}

\caption{ $\Delta I_e(E_1,E_2,\theta)$ as a function of the phase
  $\delta$ for $E_1=400$ MeV, $E_2=1000$ MeV and several chosen values
  of azimuthal angle: $\theta=0.12\pi$ (blue dashed line),
  $\theta=0.18\pi$ (orange dotted line), $\theta=0.25\pi$ (green solid
  line) and $\theta=0.39\pi$ (red dot-dashed line).  Normal mass
  ordering is assumed.  \label{fig:deltap21}}

\end{figure}

To illustrate the dependence of $\Delta I_e(E_1,E_2,\theta)$ on the
phase $\delta$, in Fig.~\ref{fig:deltap21} we plot it for chosen
values of energy and azimuthal angles.  As expected, the dependence on
$\delta$ resembles pure sine function almost symmetric with respect to
the horizontal axis.

Achieving similar cancellation of constant term and term proportional
to $\sin^2\alpha_X$ in $I_\mu$ is also possible, but requires more
theoretical input, as the constant term depends in this case on
$\phi_X$. Using approximation of eq.~(\ref{eq:phialexp}) for $\phi_X$,
quantity $\Delta I_\mu$ defined as
\bea
\Delta I_\mu(E_1,E_2,\theta) &=& \frac{E_1^2}{E_2^2} I_\mu(E_1,\theta)
- I_\mu(E_2,\theta) - \left(\cos^4 \theta_{23} + (\cos^4 \theta_{13} +
\sin^4 \theta_{13}) \sin^4 \theta_{23} \right.\nnb\\
&+&\left. \frac{1}{4} \sin^2 2 \theta_{13} \sin^2 \theta_{23} +
\frac{1}{2} \sin^2 \theta_{13} \sin^2 2 \theta_{23} \cos 2 \phi_X
\right) \left( \frac{E_1^2}{E_2^2} -1 \right)
\label{eq:deltapmu21}
\eea
can be used to set bounds on different combination of $\sin\delta$ and
$\cos\delta$ than the one derived from $\Delta I_e$.

\section{Measurements with the finite energy and angular resolution}
\label{sec:finbin}
  
For the neutrino oscillation probabilities one can exploit the simple
energy dependence of the effective $\phi_X,\sin\alpha_X$ parameters to
simplify expressions for averaging over the finite energy bins.  Using
the approximate explicit energy scaling properties holding well in the
energy range above $300-400$ MeV:
\bea
\sin\alpha_X(E',\theta) &\approx& \frac{E}{E'} \sin\alpha_X(E,\theta)\nnb\\
\sin2 \alpha_X(E',\theta) &\approx& 2 \frac{E}{E'}
\sin\alpha_X(E,\theta)\sqrt{1 - \frac{E^2}{E^{'2}}
  \sin^2\alpha_X(E,\theta)}\nnb\\[2mm]
\phi_X(E',\theta) &\approx& \phi_X(E,\theta)
\eea
we can estimate the effect of averaging over hypothetical experimental
bins.  The integral over energy in eq.~(\ref{eq:PI}) can be calculated
analytically:
\bea
\frac{1}{\Delta E } \int_{E - \frac{\Delta E}{2}}^{E + \frac{\Delta
    E}{2}} I_{\alpha\beta}(E',\theta') dE' = I_{\alpha\beta}(E,\theta') + {\cal
  O}\left(\frac{\Delta E^2}{E^2}\right)
\label{eq:pbareint}
\eea
For the energy resolutions achievable at Dune \cite{DUNE} or HyperK
\cite{HYPERK} experiments the term ${\cal O}\left(\frac{\Delta
  E^2}{E^2}\right)$ is small and can be neglected.  Thus, after
integration over the azimuthal angle all averaged transition
probabilities $\bar P_{\alpha\beta}$ can be expressed in terms of four
functions given by the integrals
\bea
\eta(E,\theta) &=& \frac{1}{\Delta \theta} \int_{\theta -
  \frac{\Delta \theta}{2}}^{\theta + \frac{\Delta \theta}{2}}
\sin^2 \alpha_X(E,\theta')d\theta'\nnb\\
\sigma(E,\theta) &=& \frac{1}{\Delta \theta} \int_{\theta -
  \frac{\Delta \theta}{2}}^{\theta + \frac{\Delta \theta}{2}} \cos^2
\alpha_X(E,\theta')\cos 2 \phi_X(\theta')d\theta'\nnb\\
\xi_1(E,\theta) &=& \frac{1}{\Delta \theta} \int_{\theta -
  \frac{\Delta \theta}{2}}^{\theta + \frac{\Delta \theta}{2}} \sin 2
\alpha_X(E,\theta') \cos \phi_X(\theta') d\theta'\nnb\\
\xi_2(E,\theta) &=& \frac{1}{\Delta \theta} \int_{\theta -
  \frac{\Delta \theta}{2}}^{\theta + \frac{\Delta \theta}{2}} \sin 2
\alpha_X(E,\theta') \sin \phi_X(\theta') d\theta'
\eea
E.g., $\bar P_{\mu e}$ reads then as
\bea
\bar P_{\mu e} &=& 2 \cos^2\theta_{13} \sin^2\theta_{13}
\sin^2\theta_{23} + \cos^2\theta_{13} ( \cos^2\theta_{23}-
\sin^2\theta_{13} \sin^2\theta_{23} )\,\eta(E,\theta) \nnb\\
&-& \frac{1}{2} \cos^2\theta_{13} \sin\theta_{13} \sin2
\theta_{23}\,(\xi_1(E,\theta)\sin\delta + \xi_2(E,\theta) \cos\delta)
\label{eq:asolvebar}
\eea
Since in the presented formalism $\alpha_X$ and $\phi_X$ are known as
regular functions of the neutrino energy and the azimuthal angle, for
any value of $E$ and $\theta$ the coefficients $\eta,\xi_1,\xi_2$ can
be easily calculated by simple 1-dimensional numerical integration.
Therefore measurements done for different angular momentum bins can
provide information on different (but known) combinations of
$\sin\delta$ and $\cos\delta$, ultimately giving a good chance to
measure the CP-phase itself.  Obviously, if necessary one can evaluate
them also assuming more complicated Earth structure models including
more internal layers, like the full PREM model~\cite{PREM}.

The same procedure can be applied to averaging of the quantities
$\Delta I_e, \Delta I_\mu$, defined in Sec.~\ref{sec:opti}.  Up to
corrections of the order of ${\cal O}\left(\frac{\Delta
  E^2}{E^2}\right)$, the same cancellations between terms as in
eq.~(\ref{eq:deltap21}) occur for barred probabilities and we can
define observable $\Delta \bar P_{e}$ averaged over the energy and
angular bin as
\bea
\Delta \bar P_e(E_1,E_2,\theta) &=& \frac{E_1^2}{E_2^2} \bar
P_{e}(E_1,\theta) - \bar P_{e}(E_2,\theta) - \left(1 - \frac{\sin^2 2
  \theta_{13} \cos 2 \theta_{23}}{2} \right)\left(
\frac{E_1^2}{E_2^2}-1\right) \nnb\\[2mm]
&\approx& \frac{1}{\Delta \theta} \int_{\theta - \frac{\Delta
    \theta}{2}}^{\theta + \frac{\Delta \theta}{2}} \Delta
I_e(E_1,E_2,\theta') d\theta'\\[2mm]
&=& - \cos^2\theta_{13} \sin\theta_{13} \sin 2 \theta_{23}
\left(\rho_1(E_1,E_2,\theta) \sin\delta + \rho_2(E_1,E_2,\theta)
\cos\delta\right)\nnb
\label{eq:dpbardef}
\eea
where the functions $\rho_1(E_1,E_2,\theta),\rho_2(E_1,E_2,\theta)$
are defined as
\bea
\rho_1(E_1,E_2,\theta) &=& \frac{E_1^2}{E_2^2}\xi_1(E_1,\theta) -
\xi_1(E_2,\theta)\nnb\\[2mm]
\rho_2(E_1,E_2,\theta) &=& \frac{E_1^2}{E_2^2}\xi_2(E_1,\theta) -
\xi_2(E_2,\theta)
\label{eq:abfun}
\eea
In a similar manner, $\bar P_\mu$ can be expressed in terms of
$\xi_1(E,\theta)$, $\xi_2(E,\theta)$ and $\sigma(E,\theta)$. For the
antineutrino oscillation probabilities, due to their more complicated
energy dependence, averaging of $I_{\alpha\beta}$ over both energy and
angle needs to be performed using numerical integration.

\section{Summary}
\label{sec:summary}

We have investigated flavour oscillations of neutrinos and
antineutrinos created in the atmosphere by cosmic ray interactions
with the air and traversing the Earth.  We have focused on sub-GeV
neutrinos/antineutrinos ($E < {\cal O}(1)$~GeV) where CP violation
effects are large but the oscillation probabilities vary very fast
with neutrino energy and its azimuthal angle, far beyond the typical
experimental resolution.  Therefore, the "observables", carrying the
physical information, are the averaged probabilities, where the fast
oscillation pattern is averaged out.  Using the Earth model with
layers of constant matter density, we have derived very simple
analytic formulae for those averaged probabilities.  There are three
main formulae summarising our results.  Equation~(\ref{eq:PAB}) is the
most general expression suitable for fast numerical calculations of
the oscillations probabilities averaged over any experimental bins
larges than the oscillation periods.
Equations~(\ref{eq:PI}--\ref{eq:asolve}) give very accurate
approximation to the averaged probabilities, where all matter effects
are encoded in two effective parameters.  And finally,
eqs.~(\ref{eq:phialexp},\ref{eq:nuconst},\ref{eq:antiphialexp},\ref{eq:antinudef})
provide for the neutrino/antineutrino energies larger than $300-400$
MeV approximate simple analytical expressions for these effective
parameters.

The obtained analytical parametrization is very accurate when compared
with the exact numerical calculations.  It opens up the possibility of
better understanding the dependence of the averaged flavour
oscillations of sub-GeV atmospheric neutrinos as a function of their
energy and the azimuthal angle with which they hit the detector.  In
turn, our results can be useful in optimising the experimental
measurements of the leptonic CP phase in oscillations of sub-GeV
atmospheric neutrinos.  We have made several suggestions in that
direction, such as the best choice of the azimuthal angles or taking
combinations of the data that are directly measuring the CP phase.

\section*{Acknowledgements}

The work of SP is supported in part by the Polish National Science
Centre under the Beethoven series grant number
DEC-2016/23/G/ST2/04301.  The work of JR is supported in part by the
Polish National Science Centre under the grant number
DEC-2019/35/B/ST2/02008.  AI would like to thank support from the COST
Action CA18108. JR would also like to thank CERN for hospitality
during his visits there.

\newpage

\appendix

\noindent {\Large \bf Appendix}

\section{Oscillation lengths}.
\label{app:trig}

For completeness we include expressions for the length of the neutrino
tracks in Earth layers and in the atmosphere.   We consider the latter
because despite the fact that neutrinos passing through the atmosphere
only do not have time to oscillate, they can be important for
azimuthal angle $\theta\approx\pi/2$ since they come from full 360
degree plane, while those passing through Earth core come only from
the small cone.   Thus atmospheric-only neutrinos may produce serious
background.

Calculating track lengths is a straightforward exercise in
trigonometry.  We assume setup defined in Fig.~\ref{fig:earth}, with
detector at distance $h$ below Earth surface (it is 1600m for Dune and
650m for HyperK) and atmosphere width denoted by $a$.   Obviously
$h,a\ll r_i,R$ thus, in all expressions below we neglect quadratic
terms $h^2, a^2$.   Let's consider 3 cases:

1) Neutrino track length in the atmosphere.
\bea
\Delta x_{atm} = a\, |\cos\theta|\left( 1 + \frac{2 \tan^2\theta}{\sqrt{1 +
    \frac{2(a+h)}{R} \tan^2\theta } + \sqrt{1 + \frac{2
      h}{R}\tan^2\theta } }\right) \qquad 0\leq \theta \leq \pi
\eea

%
%
%
%

2) Neutrino track length in the most outer layer (``crust'').

\bigskip

Let's define $\theta_{li}$ as angles for which neutrino track is
tangent to $i$-th layer:
\bea
\sin\theta_{li} = \frac{r_i}{R-h}\qquad\qquad i=1,2,3,4
\eea
Then for $\theta \geq \theta_{l2}$ neutrino has in 1st layer single
undivided track with the length
\bea
\frac{\Delta x_1}{R} = 
\cos\theta + \sqrt{\cos^2\theta + \frac{2h}{R}} -
\frac{h}{R}(\cos\theta + |\cos\theta|) && \theta \geq \theta_{l2}
\eea
For $\theta \leq \theta_{l2}$ track has 2 parts, next to detector and
on the opposite side of Earth:
\bea
\frac{\Delta x_1^{near}}{R} & = &
\left(1 - \frac{h}{R}\right)\cos\theta -  \sqrt{\frac{r_2^2}{R^2} -
  \left(1 - \frac{2h}{R}\right)\sin^2\theta}\nnb\\
\frac{\Delta x_1^{far}}{R} & = & \sqrt{\cos^2\theta +
  \frac{2h}{R}\sin^2\theta} - \sqrt{\frac{r_2^2}{R^2} - \left(1 -
  \frac{2h}{R}\right)\sin^2\theta} 
\eea

3) Neutrino track length in inner layers.

\bigskip

For the more compact notation denote additionally $r_6=0$ and
$\theta_{l6}=0$.  For $i=2,3,4,5$ we get again single track for
$\theta_{li} \leq\theta \leq \theta_{l(i+1)}$:
\bea
\frac{\Delta x_i}{R} = 2 \sqrt{\frac{r_{i}^2}{R^2} - \left(1 -
  \frac{2h}{R}\right)\sin^2\theta}
\eea
and 2 tracks of identical length for $\theta \geq \theta_{li}$:
\bea
\frac{\Delta x_i^{near}}{R} = \frac{\Delta x_i^{far}}{R} =
\sqrt{\frac{r_{i}^2}{R^2} - \left(1 - \frac{2h}{R}\right)\sin^2\theta}
- \sqrt{\frac{r_{i-1}^2}{R^2} - \left(1 -
  \frac{2h}{R}\right)\sin^2\theta}
\eea

\section{Quality of analytical approximations}
\label{app:quality}

\subsection{Averaged oscillation probability}
\label{app:paver}

In order to test the quality of approximation of eq.~(\ref{eq:PAB}),
we employ the following procedure.
\begin{enumerate}
  \item We numerically diagonalize Hamiltonian of
    eq.~(\ref{eq:electroweak}) in each Earth layer and calculate the
    full transition matrix without any approximations, multiplying
    layer transition matrices as in eq.~(\ref{eq:stprod}).  Resulting
    transition probability, $P(E,\theta)$ of eq.~(\ref{eq:pnonaver}),
    is exact but exhibits fast variations with neutrino energy and
    with the azimuthal angle.
  \item We average $P(E,\theta)$ over energy with the use of numerical
    integration, using the formula
\bea
\hat P(E,\theta) = \frac{1}{4 \Delta E}\int_{E - 2 \Delta E}^{E + 2
  \Delta E} P(E',\theta) dE'
\label{eq:intaver}
\eea
Averaging is done approximately over 4 periods $\Delta E$ of ``fast''
oscillations in energy, which (in vacuum) are given by
\bea
\Delta E = \frac{4 \pi E^2}{\Delta m_{a}^2 L(\theta)}
\eea
where $L(\theta)$ is the total neutrino track length in Earth for a
given azimuthal angle.  Actual value of $\Delta E$ in matter differ
from the vacuum, but tests show that the result of numerical averaging
is stable against variations of $\Delta E$ as long as it has the
correct order of magnitude and we integrate over several (here 4)
periods $\Delta E$.
\item For each Earth layer we diagonalize numerically Hamiltonian
  ${\cal H}'$ and calculate relevant transition matrix $S_i$ in
  rotated basis of eq.~(\ref{eq:rotbasis}).  We assume the upper
  $2\times 2$ sub-block of $S_i$ to be matrix $X_i$, as defined in
  eq.~(\ref{eq:prod3}).  Further, we evaluate full matrix $X$ as a
  time-ordered product of $X_i$ (see eq.~(\ref{eq:xfull})).  Finally,
  knowing matrix $X$ and hence also the matrices $A,B$ defined in
  eq.~(\ref{eq:sfin}), we calculate the quantity $I$ (see
  eq.~(\ref{eq:idef})).  Finally, for the analytically averaged
  oscillation probability we use the approximation $\bar P(E,\theta)
  \approx I(E,\theta)$, as discussed in Sec.~\ref{sec:genopti}.
\end{enumerate}

\begin{figure}[tb!]
\begin{center}
\begin{tabular}{cc}
\includegraphics[width=0.45\textwidth]{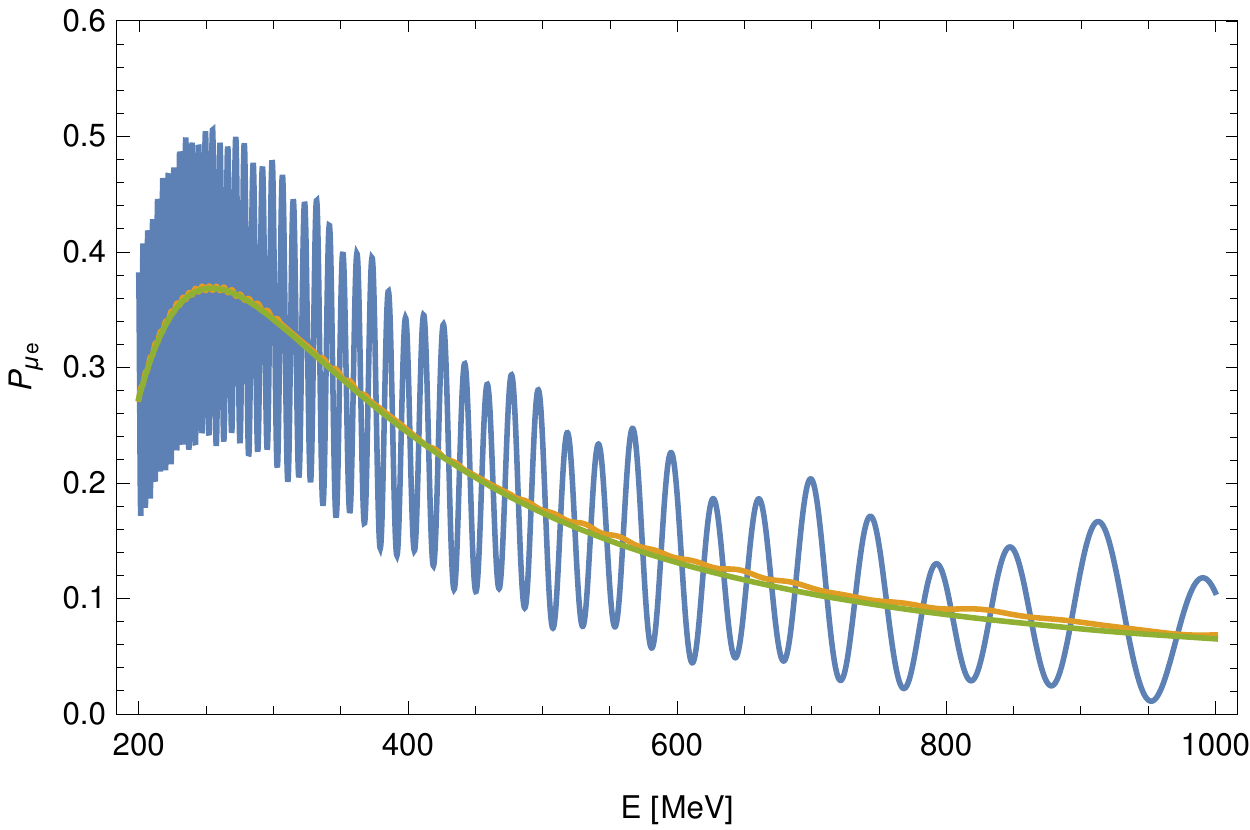} &

\includegraphics[width=0.45\textwidth]{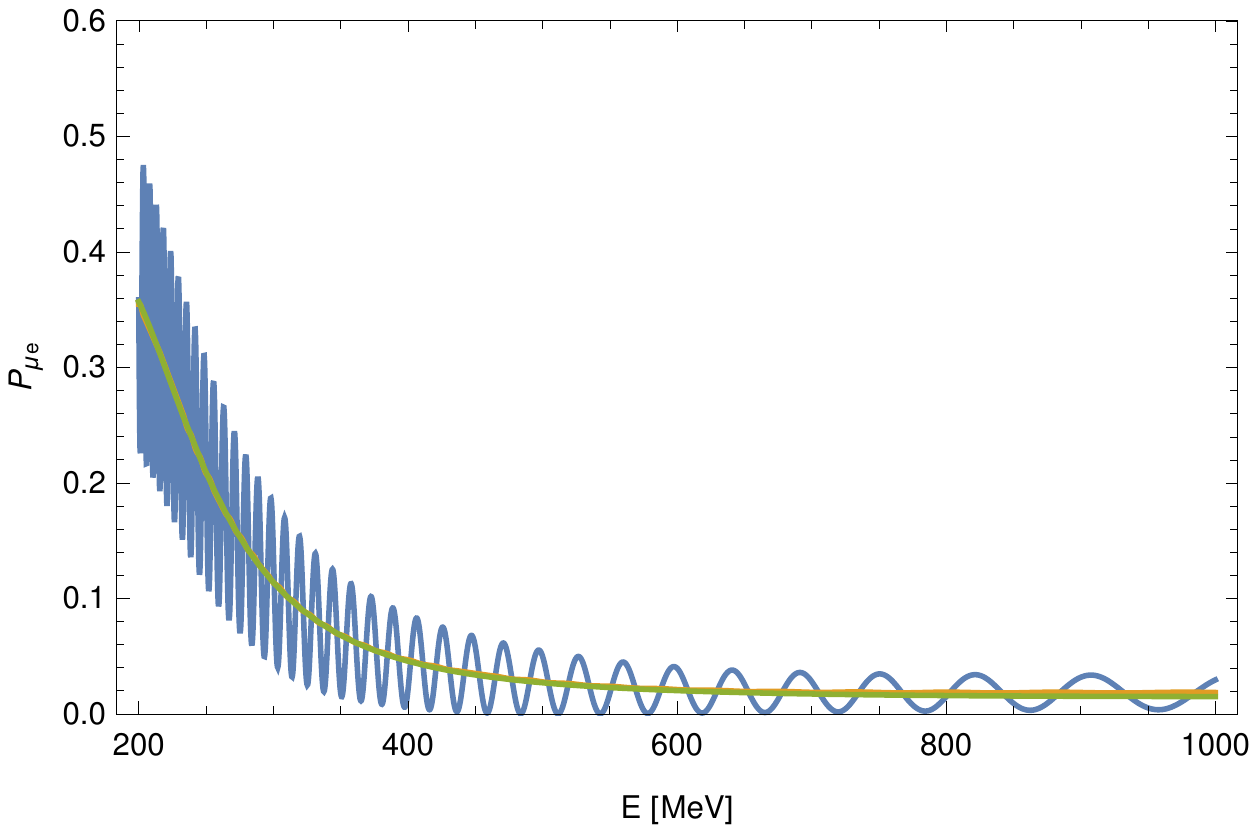}\\[3mm]
\includegraphics[width=0.45\textwidth]{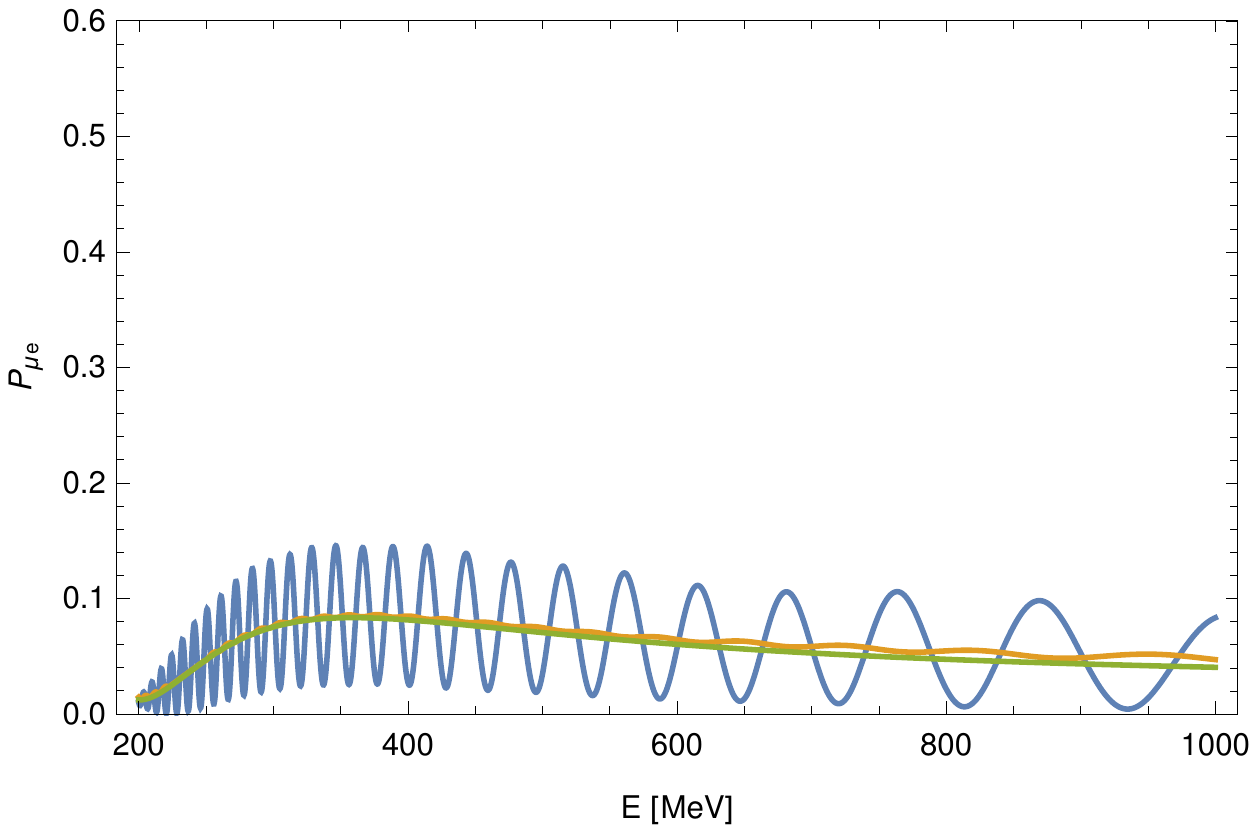} &
\includegraphics[width=0.45\textwidth]{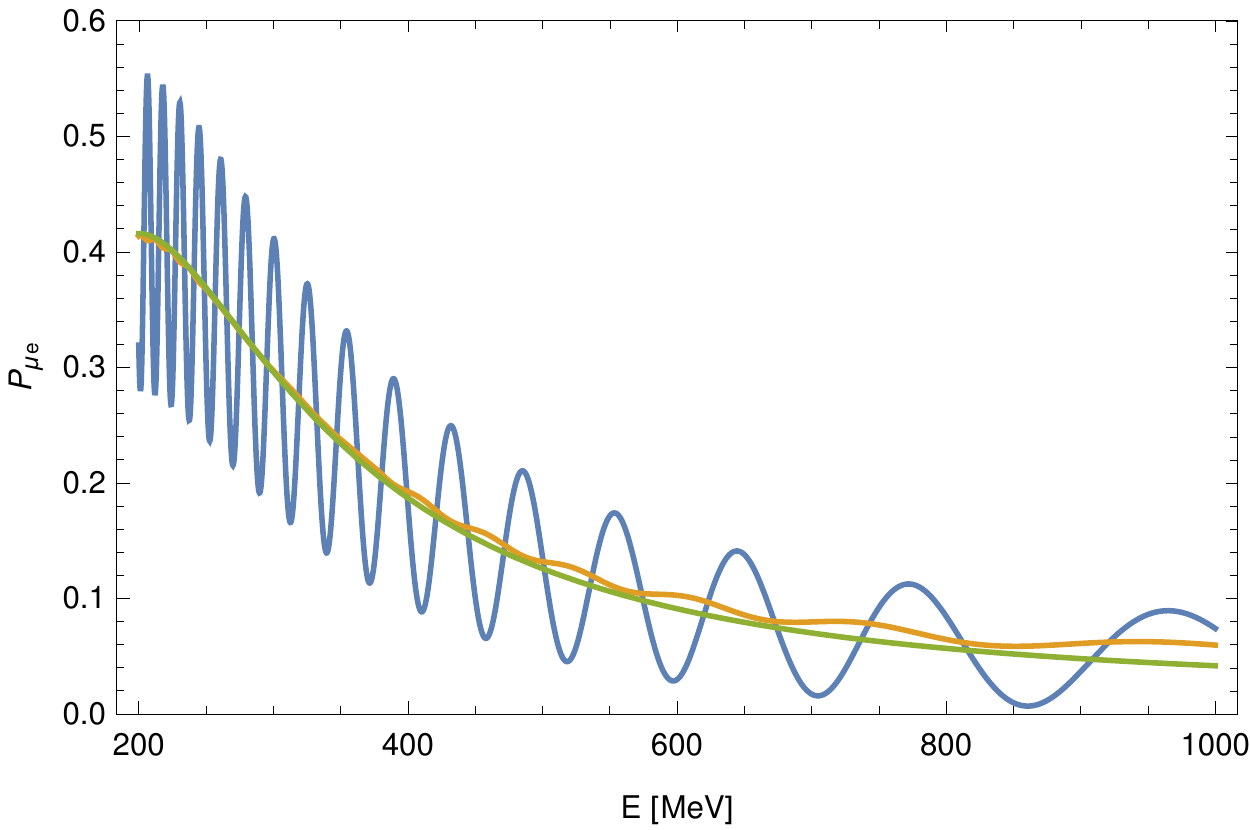}
\end{tabular}
\end{center}

\caption{Oscillation probabilities for $\nu_\mu\to \nu_e$ transitions
  for the CP-phase $\delta=\pi/2$ as a function of neutrino energy (in
  MeV) and different azimuthal angles.  Blue line: $P_{\mu
    e}(E,\theta)$, yellow line: $\hat P_{\mu e}(E,\theta)$, green
  line: $\bar P_{\mu e}(E,\theta)$.  Upper left/right panels:
  $\theta=\pi/10$ and $\theta=\pi/4$, respectively.  Lower left/right
  panels: $\theta=\pi/3$ and $\theta=\pi/2.5$.  Normal mass ordering
  is assumed.
   \label{fig:eaccu} }

\end{figure}

The comparison of $P$, $\hat P$ and $\bar P$ is illustrated in
Fig.~\ref{fig:eaccu}.  In general, analytical average of
eq.~(\ref{eq:PAB}) works very well in the sub-GeV range, some
differences between $\bar P$ and $\hat P$ can be attributed more to
the inaccuracies in numerical integration rather then in the
approximations used when deriving the formula~(\ref{eq:PAB}).

\begin{figure}[tb!]
\begin{center}
\includegraphics[width=0.6\textwidth]{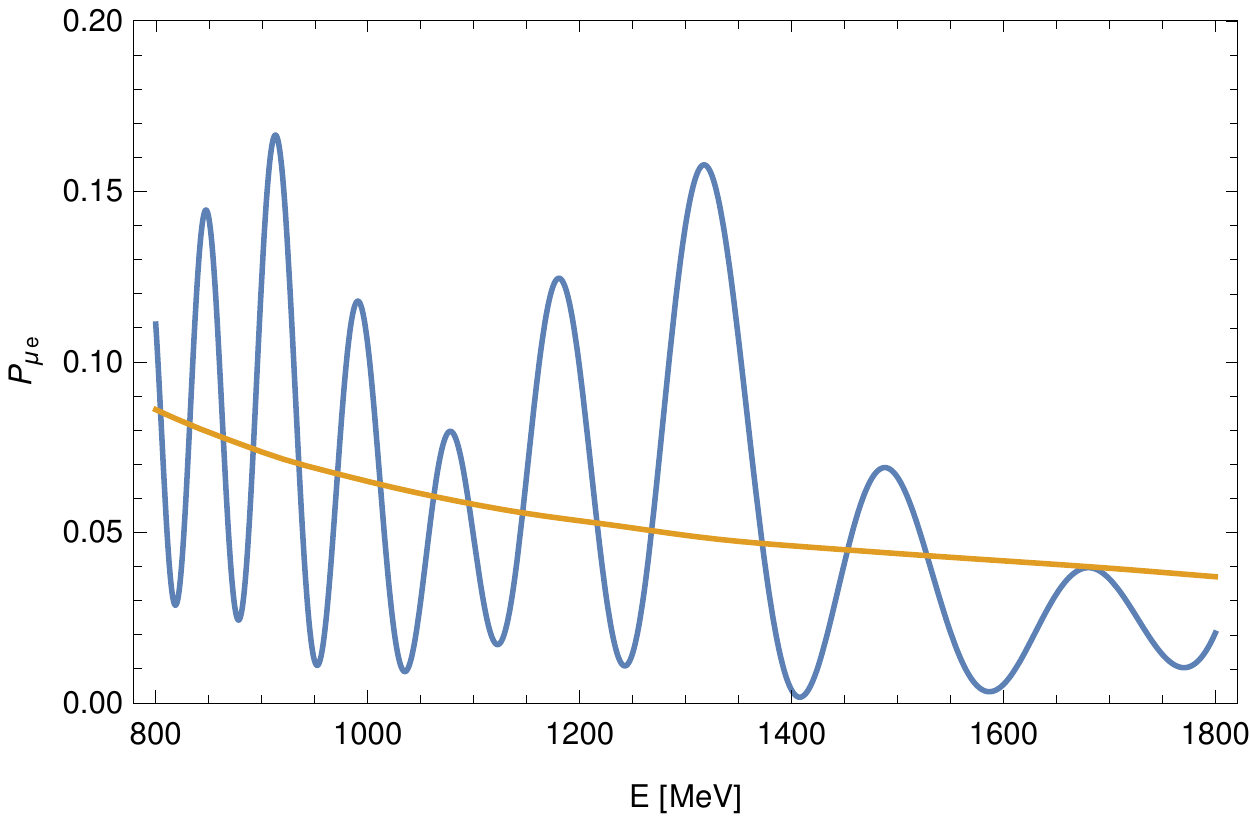} 
\end{center}

\caption{Oscillation probabilities for $\nu_\mu\to \nu_e$ transitions
  for the CP-phase $\delta=\pi/2$ as a function of neutrino energy (in
  MeV) and azimuthal angle $\theta=\pi/10$.  Blue line: $P_{\mu
    e}(E,\theta)$, green line: $\bar P_{\mu e}(E,\theta)$.  Normal
  mass ordering is assumed.  \label{fig:eacculong} }

\end{figure}

For the neutrino energies exceeding 1 GeV, the accuracy of
approximation~(\ref{eq:PAB}) breaks down, as the variation of
probabilities with energy becomes slower and less regular (see
Fig.~\ref{fig:eacculong}). In addition, periods of oscillations may
eventually become larger than the experimental resolution in energy
and azimuthal angle.  Therefore, our analytically averaged formulae
for oscillation probabilities should be used only in the sub-GeV
neutrino energy range.

\subsection{Numerical fits for  $\alpha_X(\theta,E)$ and
  $\phi_X(\theta,E)$ angles}
\label{app:numphi}

Matrix $X$ obtained numerically as a $2\times 2$ sub-block of full
$3\times 3$ transition matrix (as described in point 3 of the previous
Section) is only approximately unitary and symmetric and has
determinant slightly different from unity.  We obtain best values of
angles $\phi_X, \alpha_X$ minimising the difference between the
symmetric form on the RHS of eq.~(\ref{eq:xidef}) and the $X$ matrix
derived by the numerical diagonalization (denoted below as $X^{num}$),
i.e.  we seek the minimum of the function
\bea
f(\alpha,\phi) &=& |X_{11}^{num} - e^{-i\phi}\cos\alpha|^2 +
|X_{22}^{num} - e^{i\phi}\cos\alpha|^2\nnb\\
&+& |X_{12}^{num} +
i\sin\alpha|^2 + |X_{21}^{num} + i\sin\alpha|^2
\eea

\begin{figure}[tb!]
\begin{center}
\includegraphics[width=0.7\textwidth]{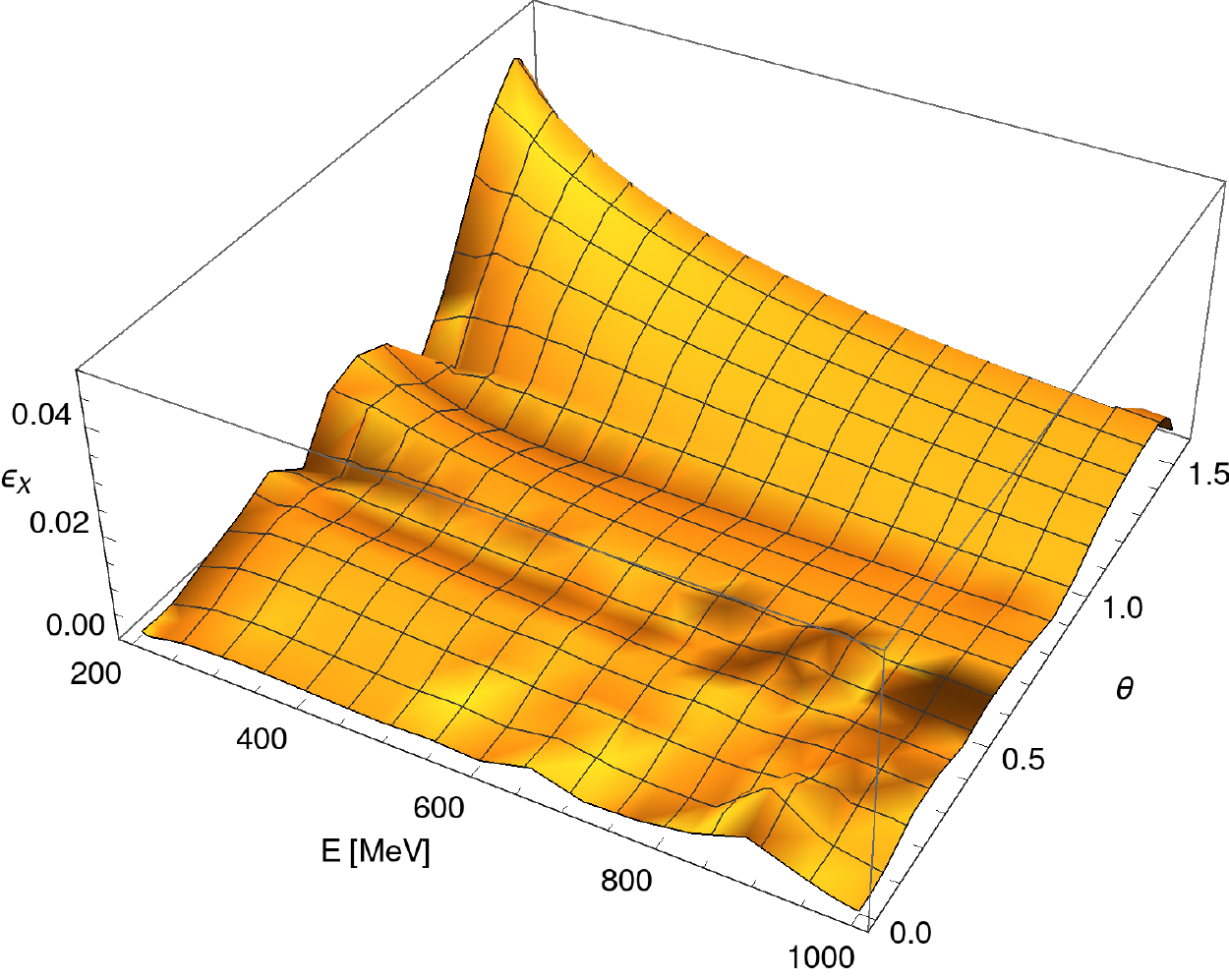}
\end{center}

\caption{ Relative error $\varepsilon_X$ of $\alpha_X,\phi_X$ fit
  plotted as a function of energy varied from 200 to 1000 MeV and
  azimuthal angle varied from 0 to $\pi/2$.  \label{fig:xaccu}}

\end{figure}

Minimisation leads to:
\bea
\phi_X &=& - \arctan \frac{\mathrm{Im}(X_{11}^{num} +
  (X_{22}^{num})^*)}{\mathrm{Re}(X_{11}^{num} + (X_{22}^{num})^*)}\nnb\\
\alpha_X &=& - \arctan \frac{\mathrm{Im}(X_{12}^{num} +
  X_{12}^{num})}{|X_{11}^{num} + (X_{22}^{num})^*|}
\label{eq:xfit}
\eea
Such a procedure reproduces very well $X$ matrix derived by numerical
diagonalization.  Fig.~\ref{fig:xaccu} shows the relative error of a
fit as a function of $E$ and $\theta$.  The error is defined as
\bea
\varepsilon_X = \frac{||X^{num} - X(\alpha_X,\phi_X)||}{||X^{num}||}
\eea
with $X(\alpha_X,\phi_X)$ defined in eq.~(\ref{eq:xfull}).

As one can see, only for low energies $E<400$ MeV and azimuthal angle
close to $\pi/2$, where the asymmetry of underground detector position
and neutrino track in atmosphere becomes relevant, the error can reach
3-4\%.  For smaller $\theta$ angles it is always small, confirming the
assumed analytical symmetry properties of $X$ matrix and justifying
the approximations done in derivation of eq.~(\ref{eq:xfull}).

\bibliography{iprr}

\providecommand{\href}[2]{#2}\begingroup\raggedright\begin{thebibliography}{10}

\bibitem{Donini:1999}
A.~Donini, M.~B. Gavela, P.~Hernandez, and S.~Rigolin, {\it {Neutrino mixing
  and CP-violation}},  {\em Nucl. Phys.} {\bf B574} (2000), no.~1-2 23--42,
  [\href{http://arxiv.org/abs/hep-ph/990}{{\tt hep-ph/990}}].

\bibitem{Ohlsson:1999um}
T.~Ohlsson and H.~Snellman, {\it {Neutrino oscillations with three flavors in
  matter: Applications to neutrinos traversing the Earth}},  {\em Phys. Lett.
  B} {\bf 474} (2000) 153--162,
  [\href{http://arxiv.org/abs/hep-ph/9912295}{{\tt hep-ph/9912295}}]. [Erratum:
  Phys.Lett.B 480, 419--419 (2000)].

\bibitem{Farzan:2002ct}
Y.~Farzan and A.~Smirnov, {\it {Leptonic unitarity triangle and CP violation}},
   {\em Phys. Rev. D} {\bf 65} (2002) 113001,
  [\href{http://arxiv.org/abs/hep-ph/0201105}{{\tt hep-ph/0201105}}].

\bibitem{Nunokawa:2007}
H.~Nunokawa, S.~J. Parke, and J.~W.~F. Valle, {\it {CP Violation and Neutrino
  Oscillations}},  {\em Prog. Part. Nucl. Phys.} {\bf 60} (2007), no.~02
  338--402, [\href{http://arxiv.org/abs/0710.0554}{{\tt arXiv:0710.0554}}].

\bibitem{Akhmedov:2008}
E.~K. Akhmedov, M.~Maltoni, and A.~Y. Smirnov, {\it {Neutrino oscillograms of
  the Earth: effects of 1-2 mixing and CP-violation}},  {\em JHEP} {\bf 06}
  (2008) 072, [\href{http://arxiv.org/abs/0804.1466}{{\tt arXiv:0804.1466}}].

\bibitem{Branco:2012}
G.~C. Branco, R.~Gonzalez~Felipe, and F.~R. Joaquim, {\it {Leptonic CP
  violation}},  {\em Rev. Mod. Phys.} {\bf 84} (2012), no.~2 515,
  [\href{http://arxiv.org/abs/1111.5332}{{\tt arXiv:1111.5332}}].

\bibitem{Ohlsson:2013}
T.~Ohlsson, H.~Zhang, and S.~Zhou, {\it {Probing the leptonic Dirac
  CP-violating phase in neutrino oscillation experiments}},  {\em Phys. Rev.}
  {\bf D87} (2013), no.~05 053006, [\href{http://arxiv.org/abs/1301.4333}{{\tt
  arXiv:1301.4333}}].

\bibitem{Razzaque:2014}
S.~Razzaque and A.~Y. Smirnov, {\it {Super-PINGU for measurement of the
  leptonic CP-phase with atmospheric neutrinos}},  {\em JEHP} {\bf 05} (2015)
  139, [\href{http://arxiv.org/abs/1406.1407}{{\tt arXiv:1406.1407}}].

\bibitem{Machado:2014}
P.~A.~N. Machado, H.~Minakata, H.~Nunokawa, and R.~Zukanovich~Funchal, {\it
  {What can we learn about the lepton CP phase in the next 10 years?}},  {\em
  JHEP} {\bf 05} (2014) 109, [\href{http://arxiv.org/abs/1307.3248}{{\tt
  arXiv:1307.3248}}].

\bibitem{Bernabeu:2018}
J.~Bernabeu and A.~Segarra, {\it {Disentangling genuine from matter-induced CP
  violation in neutrino oscillations}},  {\em Phys. Rev. Lett.} {\bf 121}
  (2018), no.~21 211802, [\href{http://arxiv.org/abs/1806.07694}{{\tt
  arXiv:1806.07694}}].

\bibitem{Kelly:2019itm}
K.~J. Kelly, P.~A.~N. Machado, I.~Martinez-Soler, S.~J. Parke, and Y.~F.
  Perez-Gonzalez, {\it {Sub-GeV Atmospheric Neutrinos and CP-Violation in
  DUNE}},  {\em Phys. Rev. Lett.} {\bf 123} (2019), no.~08 081801,
  [\href{http://arxiv.org/abs/1904.02751}{{\tt arXiv:1904.02751}}].

\bibitem{Barger:1980}
V.~Barger, K.~Whisnant, S.~Pakvasa, and P.~R.~J. N, {\it {Matter effects on
  three-neutrino oscillations}},  {\em Phys. Rev.} {\bf D22} (1980), no.~11
  2718.

\bibitem{IoaDune:2018}
A.~Ioannisian, {\it {DUNE collaboration week, CERN Jan.28-Feb.1, 2019 and DUNE
  WG meeting, October 2018}},
  \href{http://arxiv.org/abs/indico.fnal.gov/event/18736/contributions/48808/attachments
  /30464/37472/AraATM.pdf}{{\tt
  indico.fnal.gov/event/18736/contributions/48808/attachments
  /30464/37472/AraATM.pdf}}.

\bibitem{Barger:1998}
V.~Barger, T.~J. Weiler, and Whisnant, {\it {Generalized Neutrino Mixing from
  the Atmospheric Anomaly}},  {\em Phys. Lett.} {\bf B440} (1998), no.~1-2
  1--6, [\href{http://arxiv.org/abs/hep-ph/980}{{\tt hep-ph/980}}].

\bibitem{Peres:2004}
O.~L.~G. Peres and Y.~Smirnov~A, {\it {Atmospheric neutrinos: LMA oscillations,
  Ue3 induced interference and CP-violation}},  {\em Nucl. Phys.} {\bf B680}
  (2004), no.~1-3 479--509, [\href{http://arxiv.org/abs/hep-ph/030}{{\tt
  hep-ph/030}}].

\bibitem{Friedland:2004}
A.~Friedland, C.~Lunardini, and M.~Maltoni, {\it {Atmospheric neutrinos as
  probes of neutrino-matter interactions}},  {\em Phys. Rev.} {\bf D70} (2004),
  no.~11 111301, [\href{http://arxiv.org/abs/hep-ph/040}{{\tt hep-ph/040}}].

\bibitem{Huber:2005}
P.~Huber, M.~Maltoni, and T.~Schwetz, {\it {Resolving parameter degeneracies in
  long-baseline experiments by atmospheric neutrino data}},  {\em Phys. Rev.}
  {\bf D71} (2005), no.~05 053006, [\href{http://arxiv.org/abs/hep-ph/050}{{\tt
  hep-ph/050}}].

\bibitem{Hay:2012}
E.~A. Hay and D.~C. Latimer, {\it {Implications of the Dirac CP phase upon
  parametric resonance for sub-GeV neutrinos}},  {\em Phys. Rev.} {\bf D71}
  (2005), no.~05 053006, [\href{http://arxiv.org/abs/hep-ph/050}{{\tt
  hep-ph/050}}].

\bibitem{Agarwalla:2012}
S.~K. Agarwalla, T.~Li, O.~Mena, and S.~Palomares-Ruiz, {\it {Exploring the
  Earth matter effect with atmospheric neutrinos in ice}},
  \href{http://arxiv.org/abs/1212.2238}{{\tt arXiv:1212.2238}}.

\bibitem{Blennov:2013}
M.~Blennov and A.~Y. Smirnov, {\it {Neutrino Propagation in Matter}},  {\em
  Adv. High Energy Phys.} {\bf 2013} (2013) 972485,
  [\href{http://arxiv.org/abs/1306.2903}{{\tt arXiv:1306.2903}}].

\bibitem{DUNE}
R.~Acciarri and \textit{el al.} [DUNE~Collaboration], {\it {Long-Baseline
  Neutrino Facility (LBNF) and Deep Underground Neutrino Experiment (DUNE)
  Conceptual Design Report Volume 2: The Physics Program for DUNE at LBNF}},
  \href{http://arxiv.org/abs/1512.06148}{{\tt 1512.06148}}.

\bibitem{HYPERK}
L.~Abe and \textit{el al.} [Hyper-Kamiokande Proto-Collaboration], {\it
  {Physics Potentials with the Second Hyper-Kamiokande Detectorin Korea}},
  {\em Prog Theor Exp Phys} (2018) [\href{http://arxiv.org/abs/1611.06118}{{\tt
  arXiv:1611.06118}}].

\bibitem{Wolfenstein}
L.~Wolfenstein, {\it {Neutrino oscillations in matter}},  {\em Phys. Rev.} {\bf
  D17} (1978), no.~9 2369.

\bibitem{Smirnov:1985}
S.~P. Mikheev and A.~Y. Smirnov, {\it {Resonance Amplification of Oscillations
  in Matter and Spectroscopy of Solar Neutrinos}},  {\em Sov. J. Nucl. Phys.}
  {\bf 42} (1985) 913--917.

\bibitem{Akhmedov:1988}
E.~K. Akhmedov, {\it {Neutrino oscillations in inhomogeneous matter}},  {\em
  Sov. J. Nucl. Phys.} {\bf 47} (1988) 301--302.

\bibitem{Ioannisian:2018qwl}
A.~Ioannisian and S.~Pokorski, {\it {Three neutrino oscillations in matter}},
  {\em Phys. Lett.} {\bf B782} (2018) 641 -- 645,
  [\href{http://arxiv.org/abs/1801.10488}{{\tt arXiv:1801.10488}}].

\bibitem{Wang:2019yfp}
X.~Wang and S.~Zhou, {\it {Analytical solutions to renormalization-group
  equations of effective neutrino masses and mixing parameters in matter}},
  {\em JHEP} {\bf 05} (2019) 035, [\href{http://arxiv.org/abs/1901.10882}{{\tt
  arXiv:1901.10882}}].

\bibitem{Wang:2019dal}
X.~Wang and S.~Zhou, {\it {On the Properties of the Effective Jarlskog
  Invariant for Three-flavor Neutrino Oscillations in Matter}},  {\em Nucl.
  Phys. B} {\bf 950} (2020) 114867,
  [\href{http://arxiv.org/abs/1908.07304}{{\tt arXiv:1908.07304}}].

\bibitem{PREM}
A.~M. Dziewonski and D.~L. Anderson, {\it {Preliminary reference Earth model}},
   {\em Phys. of the Earth and Planetary Interiors} {\bf 25} (1981), no.~4 297
  -- 356.

\bibitem{deSalas:2017kay}
P.~F. de~Salas, D.~V. Forero, C.~A. Ternes, M.~Tortola, and J.~W.~F. Valle,
  {\it {Status of neutrino oscillations 2018: 3$\sigma$ hint for normal mass
  ordering and improved CP sensitivity}},  {\em Phys. Lett.} {\bf B782} (2018)
  633--640, [\href{http://arxiv.org/abs/1708.01186}{{\tt arXiv:1708.01186}}].

\bibitem{Indumathi:2017kxa}
D.~Indumathi, M.~Murthy, and L.~S. Mohan, {\it {Hierarchy independent
  sensitivity to leptonic $\delta_{CP}$ with atmospheric neutrinos}},  {\em
  Phys. Rev. D} {\bf 100} (2019), no.~11 115027,
  [\href{http://arxiv.org/abs/1701.08997}{{\tt arXiv:1701.08997}}].

\end{thebibliography}\endgroup
\bibliographystyle{JHEP}

\end{document}